\documentclass[useAMS,usenatbib]{mn2e}

\usepackage{longtable}
\usepackage{subfigure}
\usepackage{graphicx}

\title[DACs and dipolar magnetic fields]{Investigating the origin of cyclical wind variability in hot, 
massive stars - I. On the dipolar magnetic field hypothesis\footnotemark[1]}
\author[A. David-Uraz et al.]{A. David-Uraz,$^{1,2}$\footnotemark[2] G. A. Wade,$^{1}$ 
V. Petit,$^{3}$ A. ud-Doula,$^{4}$ J. O. Sundqvist,$^{3,5}$
\newauthor
J. Grunhut,$^{6}$ M. Shultz,$^{1,2,7}$ C. Neiner,$^{8}$ E. Alecian,$^{9,8}$ H. F. Henrichs,$^{10}$
\newauthor
J.-C. Bouret$^{11,12}$ and the MiMeS Collaboration\\ 
$^{1}$Department of Physics, Royal Military College of Canada, PO Box 17000, Stn Forces, Kingston, Canada, K7K 4B4\\
$^{2}$Department of Physics, Engineering Physics and Astronomy, Queen's University, 99 University Avenue, Kingston, Canada, K7L 3N6\\
$^{3}$Bartol Research Institute, University of Delaware, Newark, DE 19716, USA\\
$^{4}$Penn State Worthington Scranton, Dunmore, PA 18512, USA\\
$^{5}$Institut f\"{u}r Astronomie und Astrophysik der Universit\"{a}t M\"{u}nchen, Scheinerstr. 1, D-81679 M\"{u}nchen, Germany\\
$^{6}$European Organisation for Astronomical Research in the Southern Hemisphere, Karl-Schwarzschild-Str. 2, 85748,\\ Garching bei M\"{u}nchen, Germany\\
$^{7}$European Organisation for Astronomical Research in the Southern Hemisphere, Casilla 19001, Santiago 19, Chile\\
$^{8}$LESIA, UMR 8109 du CNRS, Observatoire de Paris, UPMC, Universit\'{e} Paris Diderot, 5 place Jules Janssen, F-92195\\ Meudon Cedex, France\\
$^{9}$UJF-Grenoble 1/CNRS-INSU, Institut de Plan\'{e}tologie et d'Astrophysique de Grenoble (IPAG) UMR 5274, Grenoble, F-38041, France\\
$^{10}$Astronomical Institute Anton Pannekoek, University of Amsterdam, Science Park 904, NL-1098 XH Amsterdam, the Netherlands\\
$^{11}$Laboratoire d'Astrophysique de Marseille, CNRS-Universit\'{e} de Provence, P\^{o}le de l'\'{E}toile Site de Ch\^{a}teau-Gombert, \\
38, rue Fr\'{e}d\'{e}ric Joliot-Curie 13388 Marseille cedex 13, France\\
$^{12}$NASA/Goddard Space Flight Center, Greenbelt, MD 20771 USA}
\begin{document}

\date{Submitted 20 December 2013}

\pagerange{\pageref{firstpage}--\pageref{lastpage}} \pubyear{2013}

\maketitle

\label{firstpage}

\begin{abstract}
OB stars exhibit various types of spectral variability associated with wind structures, including 
the apparently ubiquitous discrete absorption components (DACs). These are proposed to be caused by either magnetic fields or non-radial pulsations (NRPs).
In this paper, we evaluate the possible relation between large-scale, dipolar magnetic fields and
the DAC phenomenon by investigating the magnetic properties of a sample of 13 OB stars exhibiting well-documented DAC behaviour. 

Using high-precision spectropolarimetric
data acquired in part in 
the context of the Magnetism in Massive Stars (MiMeS) project, we find no evidence for surface dipolar magnetic fields in any of these stars. 
Using Bayesian inference, we compute upper limits on the strengths of the fields 
and use these limits to assess two potential mechanisms by which the field may influence
wind outflow: magnetic wind confinement and local photospheric brightness enhancements.
Within the limits we derive, both mechanisms fail
to provide a systematic process capable of producing DACs in all of the stars of our sample. Therefore, this implies that
dipolar fields are highly unlikely to be responsible 
for these structures in all massive stars, meaning that some other mechanism must come into play.
\end{abstract}


\footnotetext[1]{Based on observations collected at the Canada-France-Hawaii Telescope (CFHT) and 
T\'{e}lescope Bernard Lyot (TBL).}
\footnotetext[2]{E-mail: adavid-uraz@astro.queensu.ca}


\newpage

\section{Introduction}\label{sec:intro}

The importance of mass loss in the evolution of massive stars has been increasingly recognized over the past 20 years
(e.g. \citealt{b1}). However, the radiatively-driven winds \citep{b2} of OB stars are host to a number of forms of instability 
(e.g. \citealt{b12}) and other competing physical
processes which are not yet fully accounted for in models. Thus an important piece of the puzzle is missing
to achieve a global understanding of these stars and
of their characteristically strong outflows. This is evidenced by different forms of spectral variability in wind-sensitive lines.

First, there are stochastic variations, which can occur over very short timescales (minutes). These are believed to be related to instability mechanisms,
such as clumping, and can be found notably atop the broad emission lines of Wolf-Rayet stars (e.g. \citealt{b45}).

On the other hand, there are also cyclical (or quasi-periodic) variations which occur typically over longer timescales (for a complete review of
the various forms cyclical variations can take, see \citealt{b47}). One example consists of the so-called 
``periodic absorption modulations", or PAMs, observed in a number of OB stars (e.g. \citealt{b46}) and which manifest themselves as optical depth modulations 
in the absorption troughs of ultraviolet (UV) P Cygni profiles. They can show a ``phase-bowing", appearing at intermediate velocities and
bending slightly upwards in the dynamic spectra, therefore occuring quasi-simultaneously at all velocities shortly thereafter
(as in HD~64760, \citealt{b14}). PAM variabilities occur on intermediate timescales (hours) and their physical cause is not known. 

In parallel, one of the most common forms of cyclical variability among OB stars is the presence of 
so-called ``discrete absorption components" (DACs). These features
are formed in the UV resonance lines of hot massive stars and appear as narrow, blueward-travelling absorption structures. Their progression from
low to near-terminal velocity over time distinguishes this form of variability from the aforementioned PAMs.
As was first shown in time series of IUE spectra \citep{y1}, DACs recur cyclically on longer timescales
(days) and at relatively well-constrained periods. 
These timescales were found to be correlated with the projected rotational velocity ($v \sin i$), suggesting
that these variations are rotationally modulated \citep{b26}. 
DACs are thought to be present in all OB stars. Indeed, narrow absorption
components (NACs; narrow absorption features typically found near terminal velocity), believed to be snapshots of DACs, are 
found in nearly all massive stars observed by IUE \citep{b13}.
However, this does not mean that all DACs are identical. Their depths vary from one star to another (they can even be opaque), 
and it is possible to find more than one DAC at
a time in single observations \citep{b4}.
Because they span the full range of velocities over time, it is believed that they are caused by large-scale
azimuthal structures extending from the base of the wind all the way to its outer regions \citep{y2}. \citet{b16} showed that a perturbation in the
photosphere could lead to co-rotating interaction regions (CIRs), although the physical nature of this perturbation is not yet known. 
This model seems consistent with the DAC phenomenon and leads to promising simulated
spectral signatures. The goal of this project is to determine what physical process constitutes the origin of DACs. Obviously, there are far-reaching
implications for the general study of massive stars, since DACs are believed to be common to all OB stars.

\begin{table*}
\caption[DAC star sample]{Sample of stars used for this study; spectral types are obtained from \citet{b3} and references therein.
$N_{\textrm{obs}}$ corresponds to the total number of independent observations for each star, $\Delta t_{\mathrm{E}}$, $\Delta t_{\textrm{N}}$ 
and $\Delta T_{\textrm{max}}$ correspond respectively to the average individual total exposure time for ESPaDOnS and NARVAL, 
and the maximum time elapsed between the first and last observation of a star on any given night (N/A for stars with only one observation
per night).}\label{tab:sample}
\begin{tabular}{|r|l|l|c|c|r|r|r|}
  \hline
HD & Name & Spectral Type & $m_{V}$ & $N_{\textrm{obs}}$ & $\Delta t_{\textrm{E}}$ & $\Delta t_{\textrm{N}}$ & $\Delta T_{\textrm{max}}$\\
   &      &               &         &                    &         (s)             &         (s)             &          (d) \\
  \hline
24912  & $\xi$ Per       & O7.5 III(n)((f)) & 4.06 & 44 & 360       & $\sim1800$ & 0.186 \\
30614  & $\alpha$ Cam    & O9.5 Ia          & 4.30 & 11 & 560       & 920        & 0.037 \\
34656  &                 & O7 II(f)         & 6.80 &  1 & 2600      & -          & N/A   \\
36861  & $\lambda$ Ori A & O8 III((f))      & 3.30 & 20 & $\sim200$ & $\sim400$  & 0.039 \\
37128  & $\epsilon$ Ori  & B0 Ia            & 1.70 & 70 & 40        & $\sim160$  & 0.122 \\
47839  & 15 Mon          & O7 V((f))        & 4.64 & 16 & 640       & $\sim1600$ & 0.035 \\
64760  &                 & B0.5 Ib          & 4.23 &  9 & 440       & -          & 0.033 \\
66811  & $\zeta$ Pup     & O4 I(n)f         & 2.25 & 30 & 80        & -          & 0.078 \\
149757 & $\zeta$ Oph     & O9.5 V           & 2.58 & 65 & 100       & 180        & 0.061 \\
203064 & 68 Cyg          & O7.5 III:n((f))  & 5.04 &  8 & 980       & $\sim2000$ & 0.053 \\
209975 & 19 Cep          & O9.5 Ib          & 5.11 & 33 & 1000      & 1800       & 0.093 \\
210839 & $\lambda$ Cep   & O6 I(n)fp        & 5.08 & 26 & -         & 2640       & N/A   \\
214680 & 10 Lac          & O9 V             & 4.88 & 36 & 400       & $\sim2000$ & 0.051 \\
  \hline
\end{tabular}
\end{table*}

The two leading hypotheses to explain DACs are magnetic fields and non-radial pulsations (NRPs). 
However, both processes present a number of challenges when it comes to explaining DACs. First, based on the statistics
of the Magnetism in Massive Stars (MiMeS) survey, less than 10\% of all massive stars are inferred to harbour detectable
magnetic fields \citep{b15a}. This is obviously a problem since DACs are thought to be common to all OB stars. On the other
hand, a pulsational origin for DACs might also be problematic, since one would expect a succession of brighter and darker
areas on the photosphere, whereas \citet{b16} specifically identify bright spots as the possible cause for DACs.
Moreover, experiments with alternating bright and dark regions, meant to simulate the brightness distribution of low-order NRPs,
failed to reproduce DAC-like variations (Owocki, priv. comm.).
On the other hand, rotational modulations (RMs; analogous to PAMs) have been modelled self-consistently with a 3D radiative transfer code 
using NRPs in HD~64760 \citep{lobel}, a star possessing DACs; however, the NRPs produce the RMs, while the DACs 
are created by introducing bright spots.
Finally, DAC recurrence timescales are deemed to be incompatible with pulsational periods and it has been suggested that this 
problem can only be solved through complex mode superpositions \citep{b40}.
This paper investigates the simplest form
of the first case: that of a purely dipolar large-scale magnetic field, inclined relative to the rotation axis. 
Indeed, most massive stars are thought to produce two DACs per rotational period \citep{b4}, so this configuration
seems like a rather natural fit. Moreover, most detected magnetic fields in OB stars are essentially dipolar, and follow the oblique
rotator model \citep{b15}. This is expected, since large-scale magnetic fields in massive stars are believed to be of fossil origin, 
relaxing into a dipolar configuration \citep{z1,z2}. On the other hand, relatively weak magnetic fields, possibly below the
threshold of detection for most MiMeS observations, could still introduce a significant modulation of the winds of OB stars.

In this paper, we examine a sample of 13 stars well known to exhibit DACs. The sample is described in detail in Section~\ref{sec:samp}. 
In Section~\ref{sec:obs}, we describe
the high-resolution spectropolarimetric observations of these stars, as well as the instruments on which they were obtained.
Section~\ref{sec:lsd} outlines the least-squares deconvolution (LSD) procedure used to maximize the signal-to-noise ratio of the Stokes
V profiles to search for Zeeman signatures. 
In Section~\ref{sec:diag}, we present the various diagnostics used to perform the most precise magnetometry 
ever obtained for this class of stars. Section~\ref{sec:notes}
contains notes on individual stars, while the results are discussed and analyzed in detail in Section~\ref{sec:disc}, as well as the conclusions of this
study and pointers for future investigations.

\section{Sample}\label{sec:samp}

Thirteen OB stars (with spectral types ranging from O4 to B0.5, and luminosity classes from V to Ia, 
see Table~\ref{tab:sample}) were selected to form this sample based on 
two main criteria: documented DAC behaviour, and available high-quality data.

All stars selected for this sample are well known to exhibit the DAC phenomenon and were extensively studied as such: 9 stars were studied by 
\citet{b4}, $\zeta$ Pup was investigated by \citet{b7}, while $\zeta$ Oph was the subject of a paper by \citet{b8}. Finally, the two B supergiants
($\epsilon$ Ori and HD~64760) were
studied by \citet{b11}. 
The suspected ubiquitous nature of DACs indicates that the physical process causing them should be common
to all OB stars. Therefore, if this process involves large-scale dipolar magnetic fields, we expect to detect such fields in most
of the stars of this sample.

Furthermore, data accessibility was one of the key factors in choosing this sample. Indeed, these stars 
were selected because available archival data (high-resolution spectropolarimetry) related to the MiMeS Project allow us to conduct 
high-precision magnetic measurements and compile a self-consistent dataset.

The stellar and wind parameters of all stars in the sample are presented in Table~\ref{tab:bp}. 
For consistency with \citet{b4}, most of the values we use are taken from that paper.
Thus, for the 11 O stars, the mass-loss
rates are obtained by applying the empirical prescription of \citet{b6}, which relies on radio free-free emission and H$\alpha$ measurements using
unclumped models. As for the 2 B stars, mass-loss rates are taken from \citet{b10} (based on optical/UV spectroscopy).
Comparison of the adopted stellar and wind parameters with more modern values (e.g. \citealt{plouc1, plouc2, plouc3, plouc4})
yield only minor differences in $T_{\textrm{eff}}$ (typically about 1 kK, $\sim 5$\%), $R_*$ (a few $R_{\odot}$, $\sim 20$\%) 
and $v_{\infty}$ (essentially identical). 
For the mass-loss rates, modern values typically differ from one another 
by a factor of a few, up to a full order of magnitude, depending on each star. In general, our
values are consistent with the lower end of that range.

\section{Observations}\label{sec:obs}

The observations were obtained at the Canada-France-Hawaii Telescope (CFHT) on the
ESPaDOnS instrument, and on its sister instrument, NARVAL, installed at T\'{e}lescope Bernard Lyot (TBL). 
Some observations were obtained as part of the Large Programs (LPs) awarded to MiMeS on both instruments, while a significant part of the dataset
was obtained as part of individual PI programs (led by V. Petit, C. Neiner, E. Alecian, H. Henrichs and J.-C. Bouret). 
Both of these instruments
are high-resolution ($R \sim 65,000$) fibre-fed \'{e}chelle spectropolarimeters. Each exposure consists of 4 sub-exposures corresponding to different
angles of the Fresnel rhomb retarders, 
which are then combined in different ways to obtain both the I (unpolarized) and V (circularly polarized) Stokes parameters, as well
as two diagnostic nulls (which have the same noise level as the V spectrum, but no stellar magnetic signal, \citealt{b17}). The spectral coverage is essentially continuous
between about 360-1000 nm. 
The reduction was performed using the Libre-ESpRIT package at the telescope, and the spectra were then normalized to the continuum. Appendix A contains
a summary of all the observations.

The use of these observations marks a significant improvement in the study of the role of magnetic fields in the generation of wind variability 
because of both their high resolution and high signal-to-noise ratio (SNR). They constitute the highest-quality
dataset compiled to date for the purpose of magnetometry on OB stars. Furthermore, the extensive time coverage obtained for a number
of stars in the sample can provide extremely tight constraints on the geometry of any surface magnetic field present (see Section~\ref{sec:diag}).
In total, this dataset is constituted of 381 spectra, for an average of nearly 30 spectra per star (HD 34656 only has 1 observation,
while $\epsilon$ Ori has 70). The data were acquired between 2006 and 2013, with a typical peak SNR of over 1000 per CCD pixel at a wavelength
of around 550 nm.

\section{Least-Squares Deconvolution}\label{sec:lsd}

In order to improve the significance of potential Zeeman signatures in the Stokes V profile, indicative of the presence of a magnetic field, 
LSD \citep{b17} was used to effectively
deconvolve each spectrum to obtain a single, high-SNR line profile. This was carried out using the latest implementation of iLSD \citep{b18}.

This procedure requires the use of a specific ``line mask" for each star, 
which is a file containing all the necessary information about the lines whose signal will be added:
central wavelength, depth and Land\'{e} factor. First, to create such a file, a line list is obtained from the Vienna Atomic Line Database (VALD, \citealt{b41}),
by inputting the effective temperature of the star, and choosing a line depth threshold (0.01 in this case). Then, the information contained in the line list is used 
to create a crude preliminary mask, which can then be filtered and adjusted. This means that some lines are removed (e.g. lines which don't actually appear
in the spectra, lines heavily contaminated by telluric absorption, hydrogen lines, due to their particular shape and behaviour, as well
as lines which were blended with hydrogen lines), while the depths of the remaining lines can be adjusted to better reproduce
the star's spectrum. This procedure also ensures that uncertainties in $T_{\rm{eff}}$ have little impact on the final mask.

Several tests were made with sub-masks to determine which of the remaining lines should be included or not.
In the end, masks including helium and metallic lines were used, as the helium lines provided most of the signal and did not alter the shape
of the mean line profile
significantly (although they do introduce extra broadening). The LSD profiles were then extracted using these masks, without applying a regularisation correction
\citep{b18} since it did not yield significant gain given the already high SNR of the spectra.

\begin{figure*}
\begin{center}
\subfigure[$\xi$ Per LSD profile.]{\includegraphics[width=2.2in]{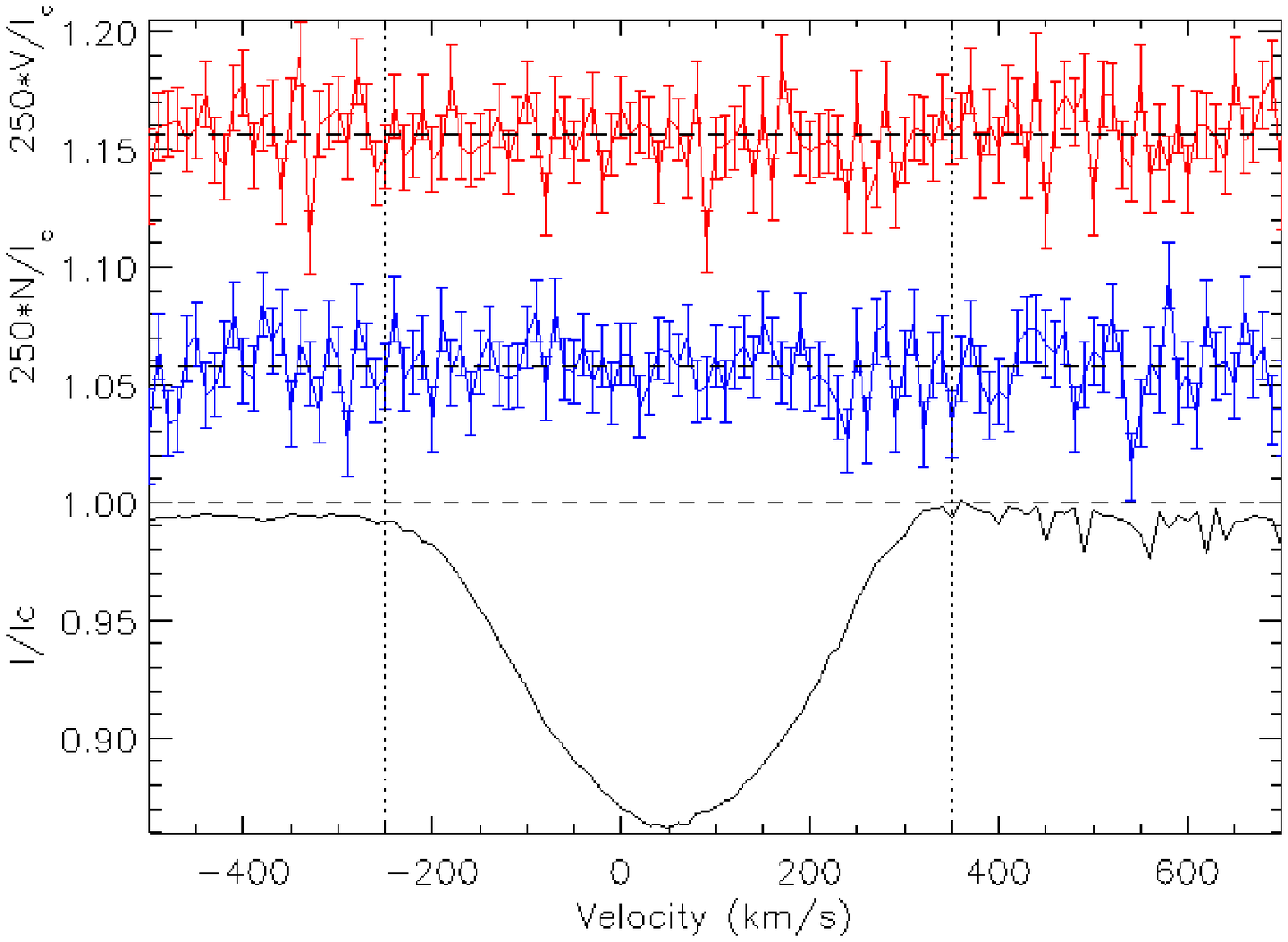}}
\subfigure[$\alpha$ Cam LSD profile.]{\includegraphics[width=2.2in]{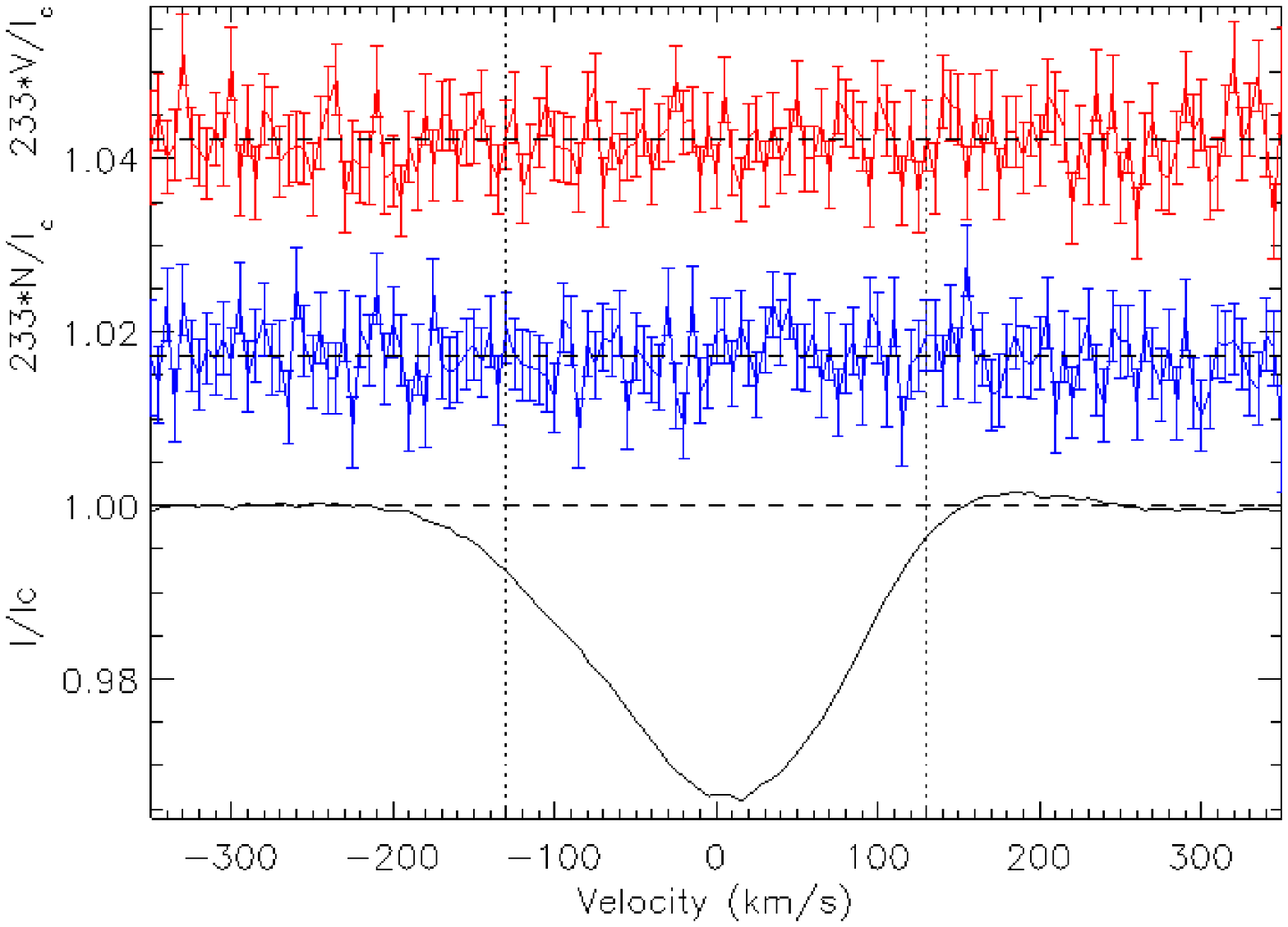}}
\subfigure[HD~34656 LSD profile.]{\includegraphics[width=2.2in]{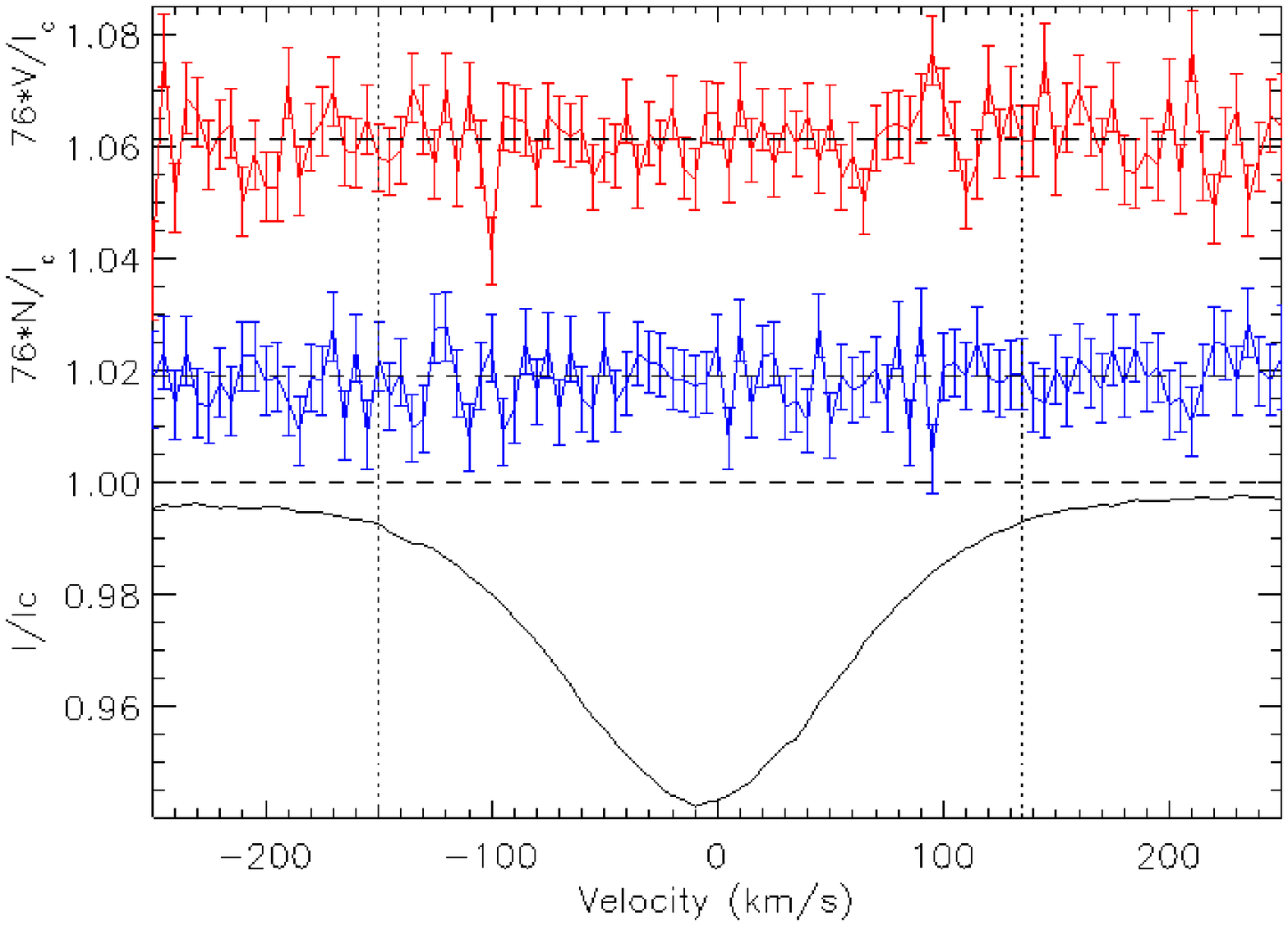}}
\subfigure[$\lambda$ Ori A LSD profile.]{\includegraphics[width=2.2in]{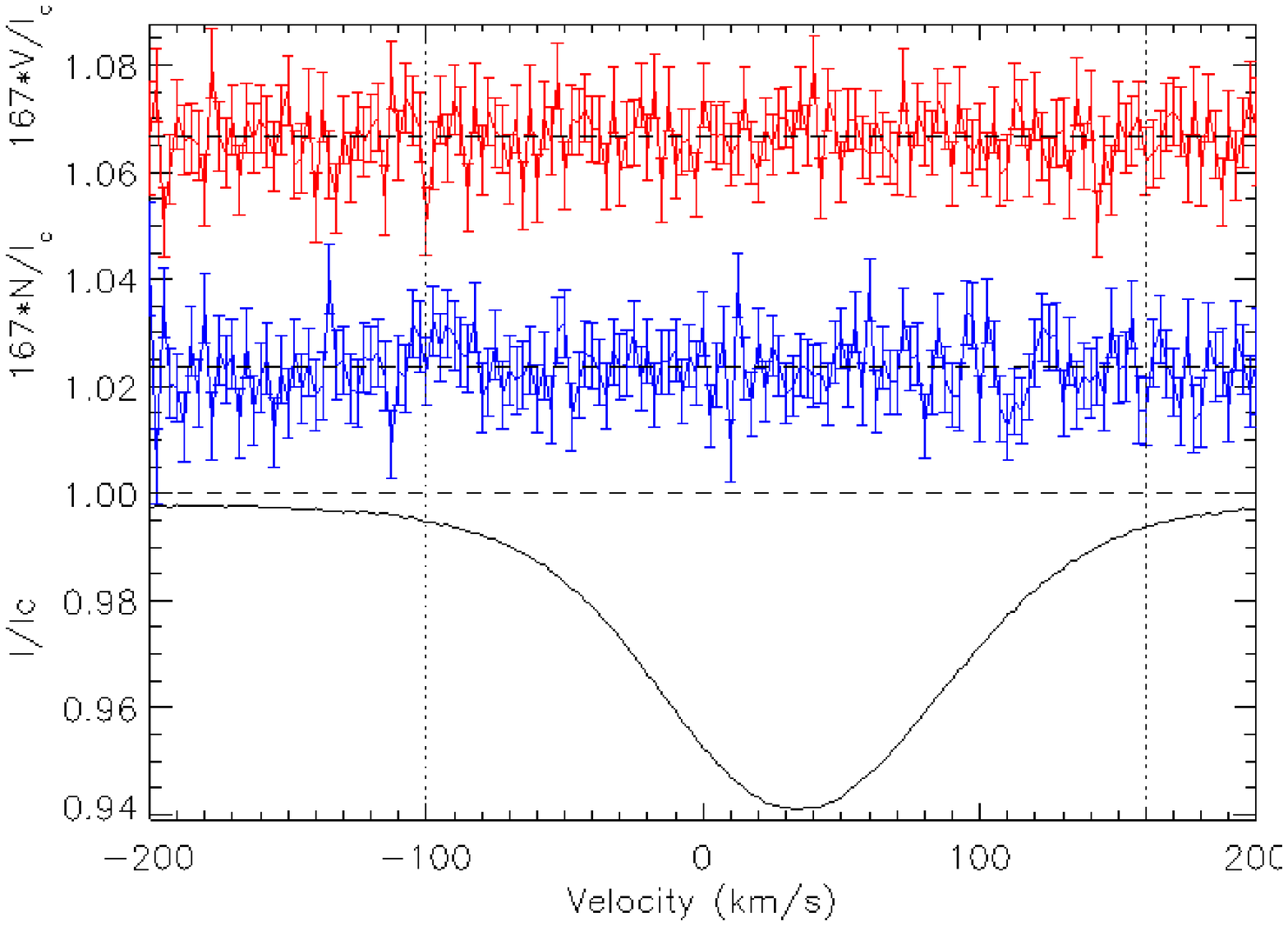}}
\subfigure[$\epsilon$ Ori LSD profile.]{\includegraphics[width=2.2in]{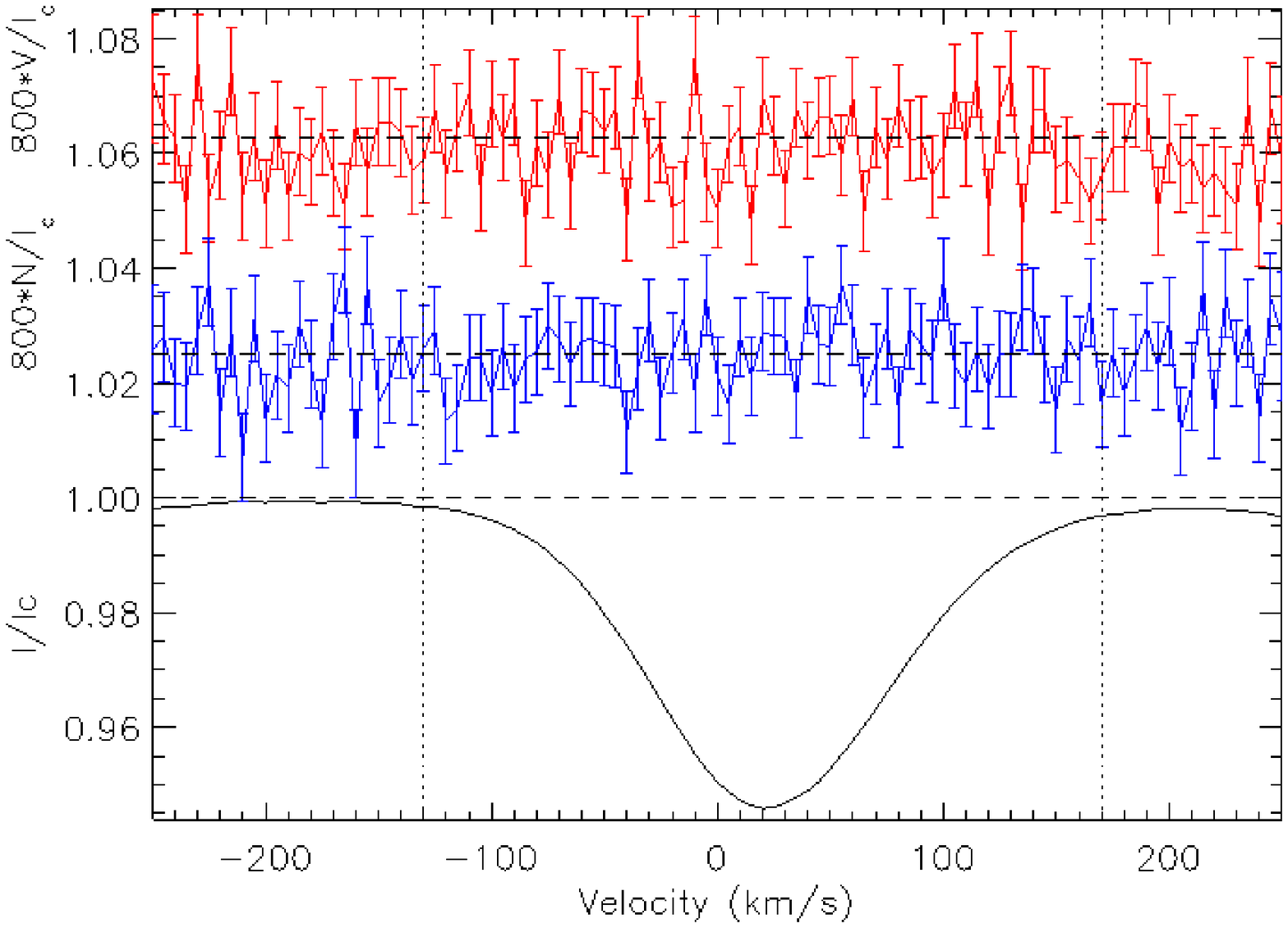}}
\subfigure[15 Mon LSD profile.]{\includegraphics[width=2.2in]{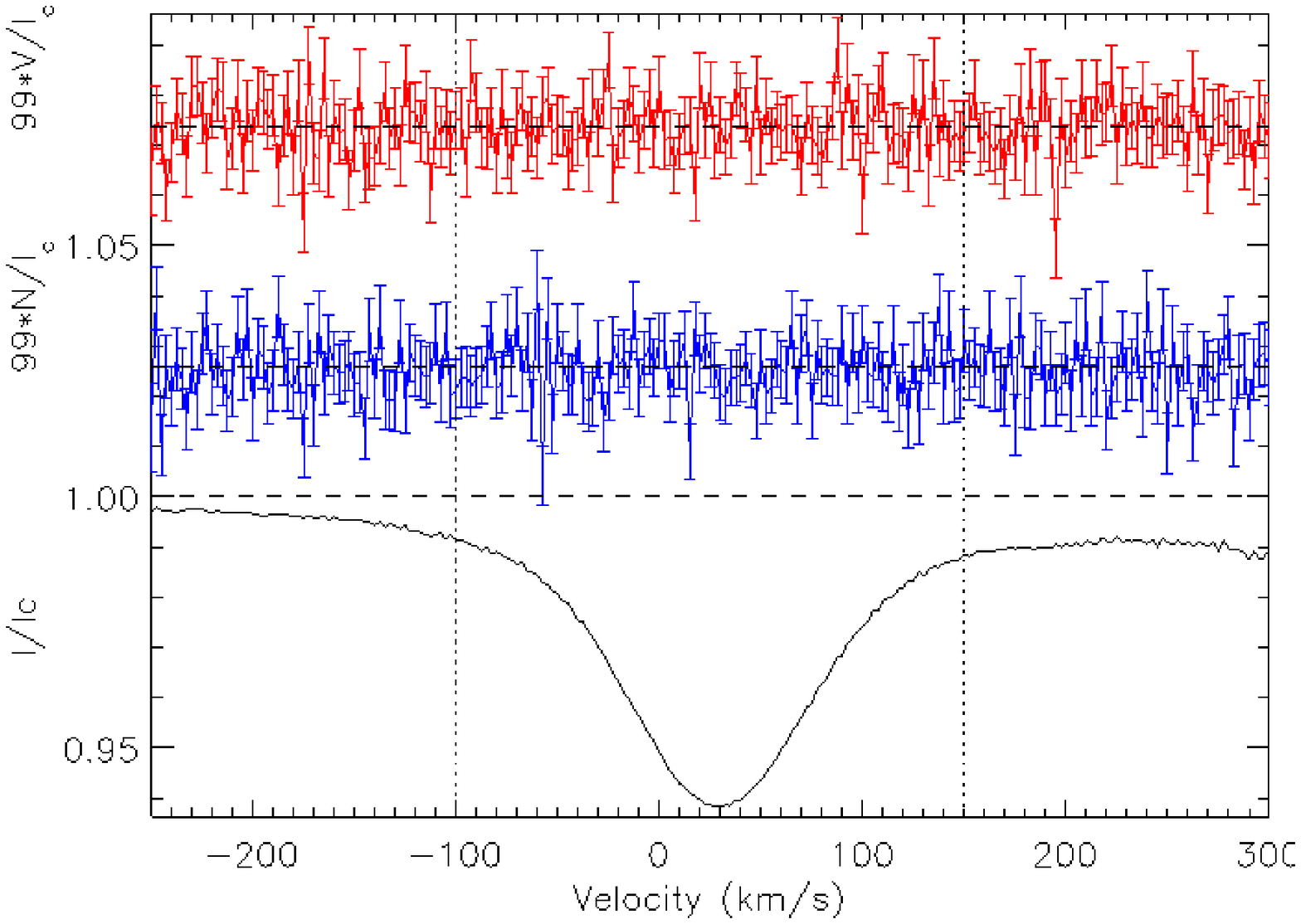}}
\subfigure[HD~64760 LSD profile.]{\includegraphics[width=2.2in]{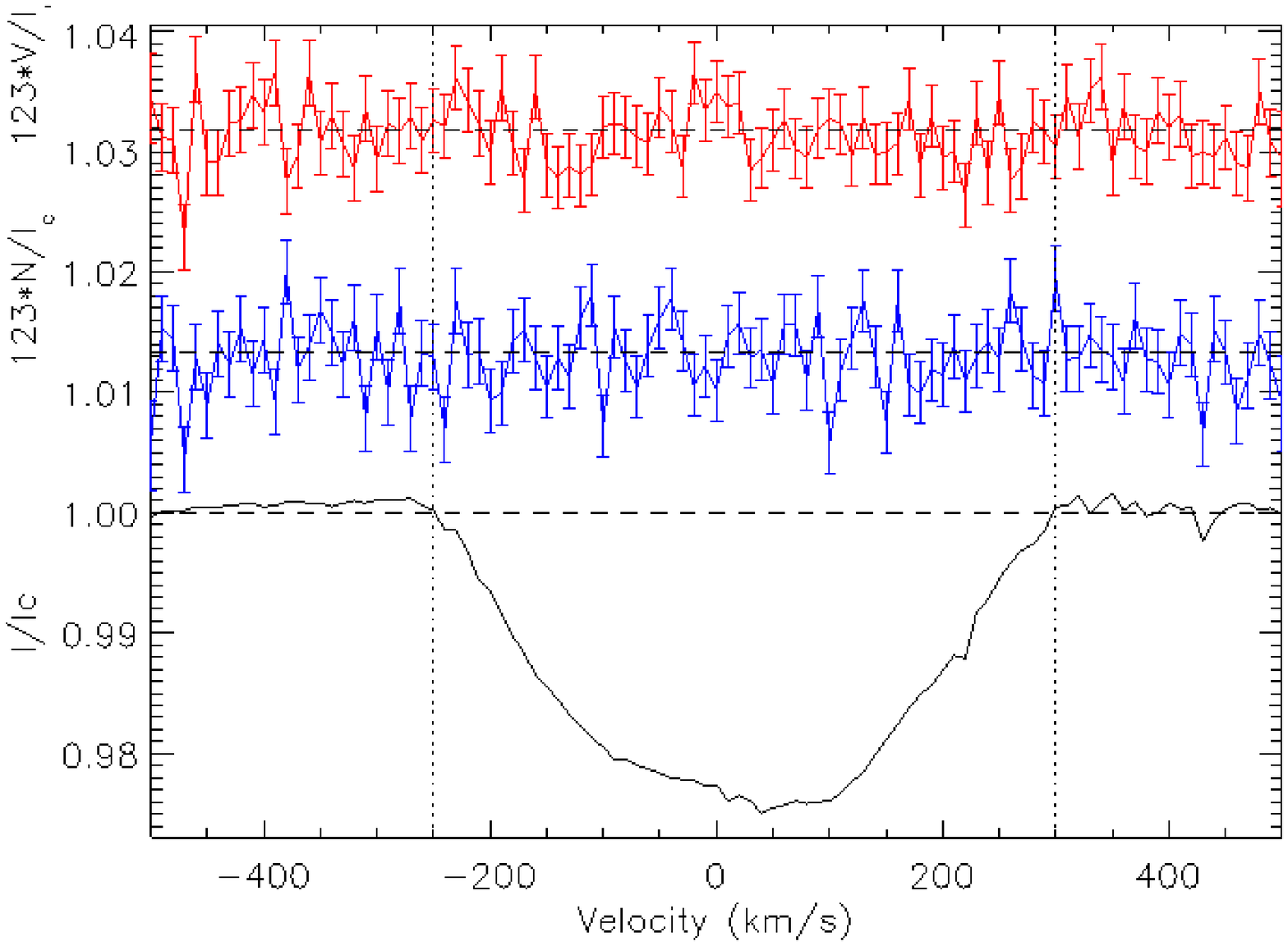}}
\subfigure[$\zeta$ Pup LSD profile.]{\includegraphics[width=2.2in]{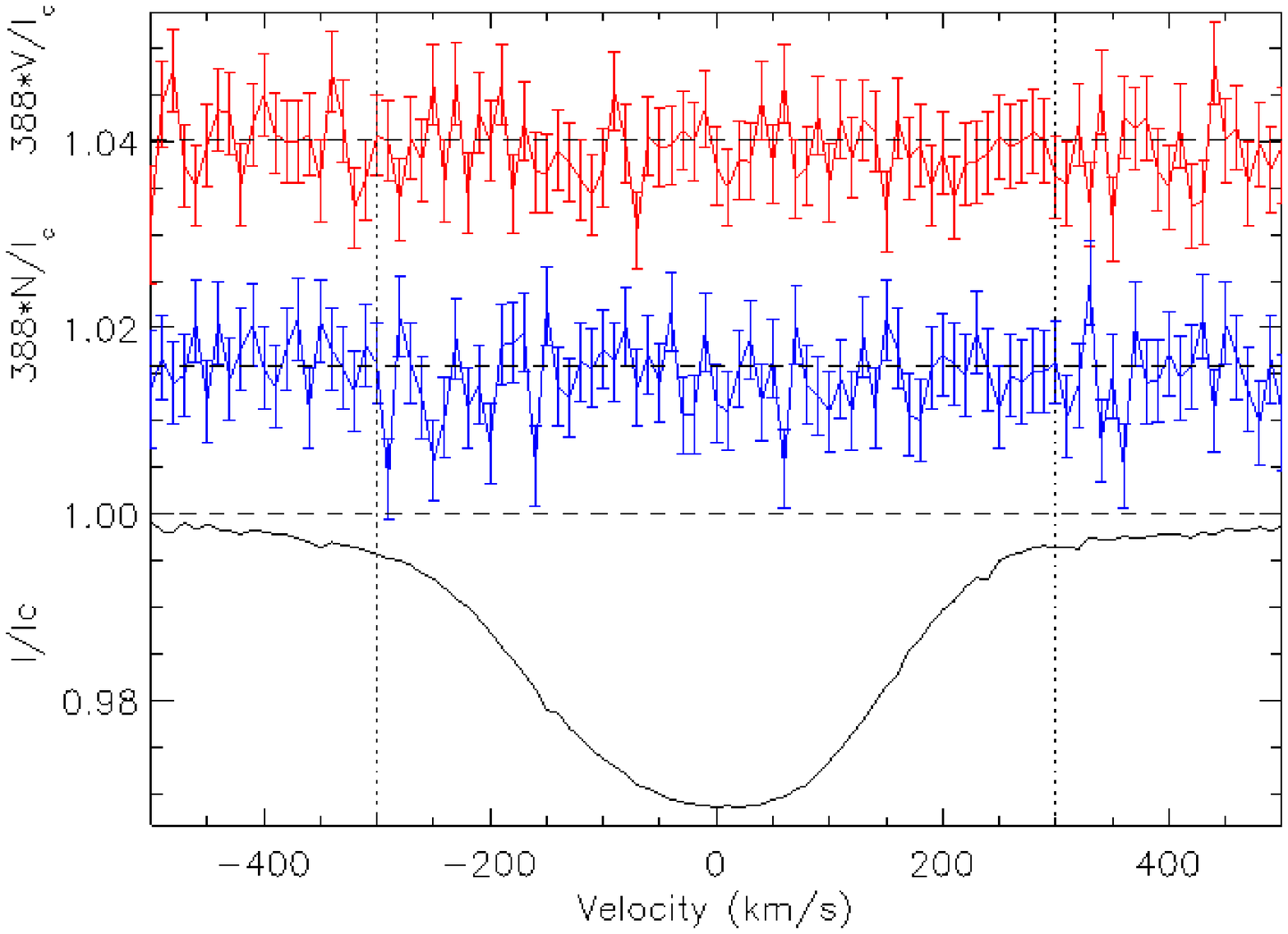}}
\subfigure[$\zeta$ Oph LSD profile.]{\includegraphics[width=2.2in]{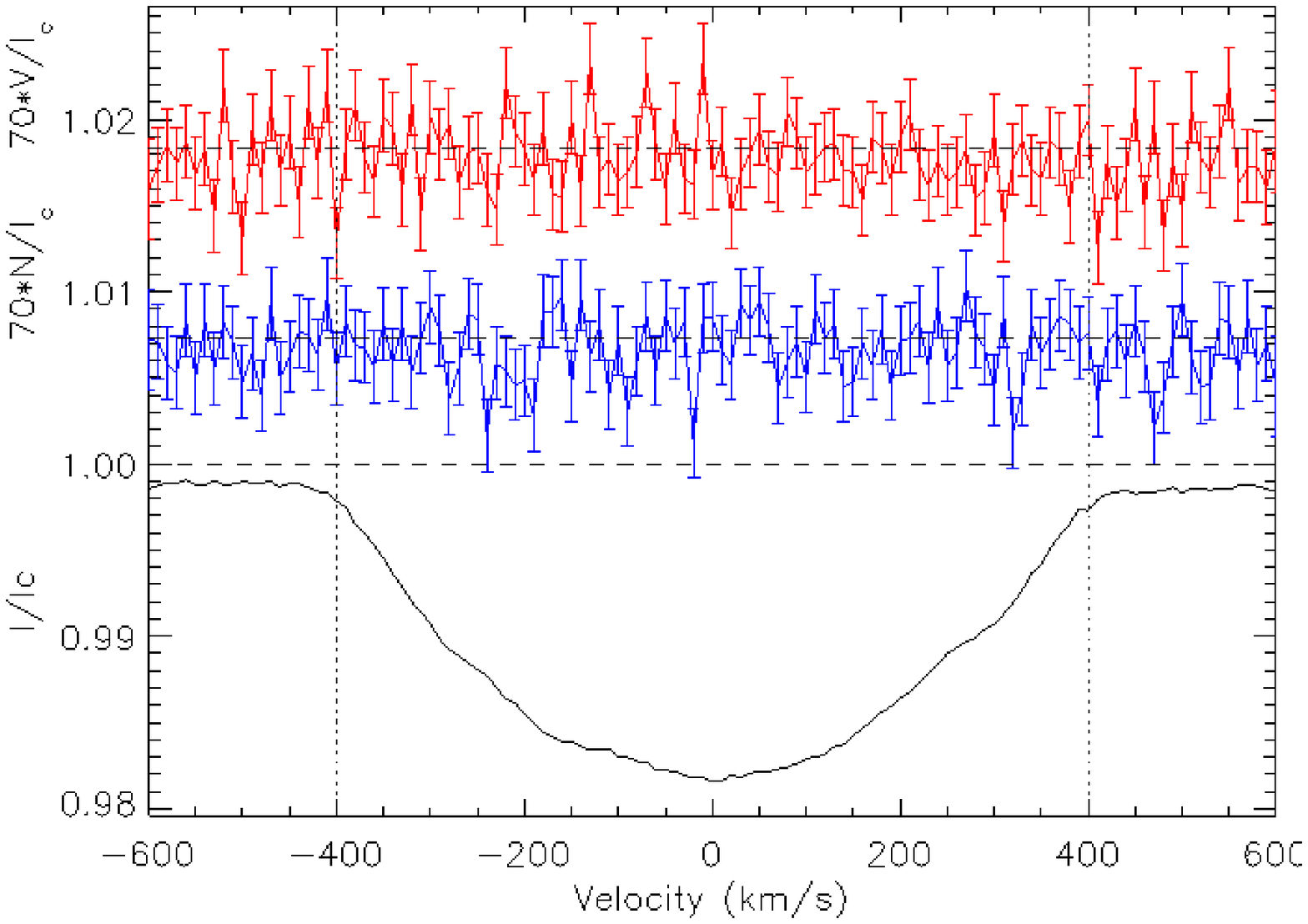}}
\subfigure[68 Cyg LSD profile.]{\includegraphics[width=2.2in]{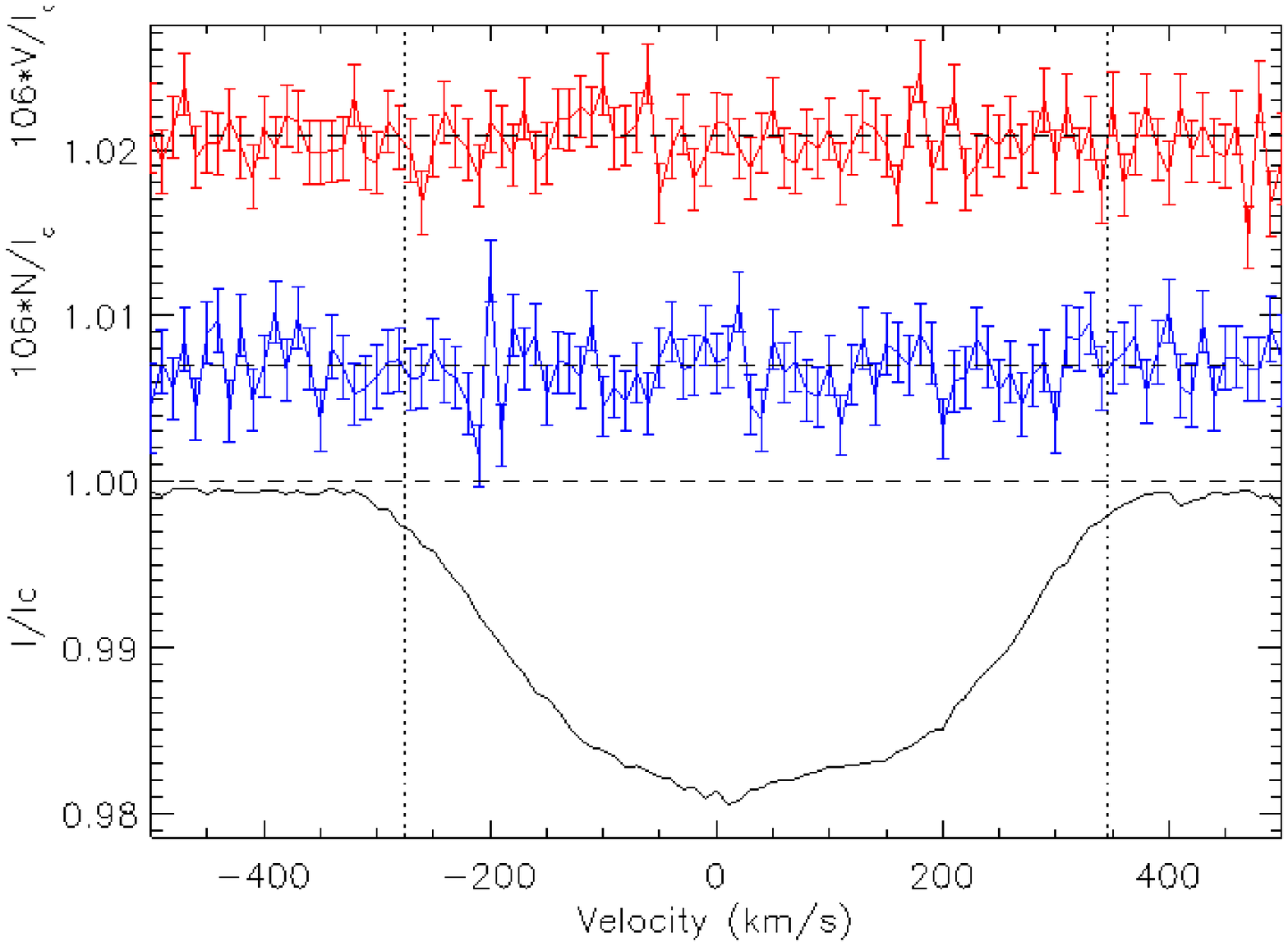}}
\subfigure[19 Cep LSD profile.]{\includegraphics[width=2.2in]{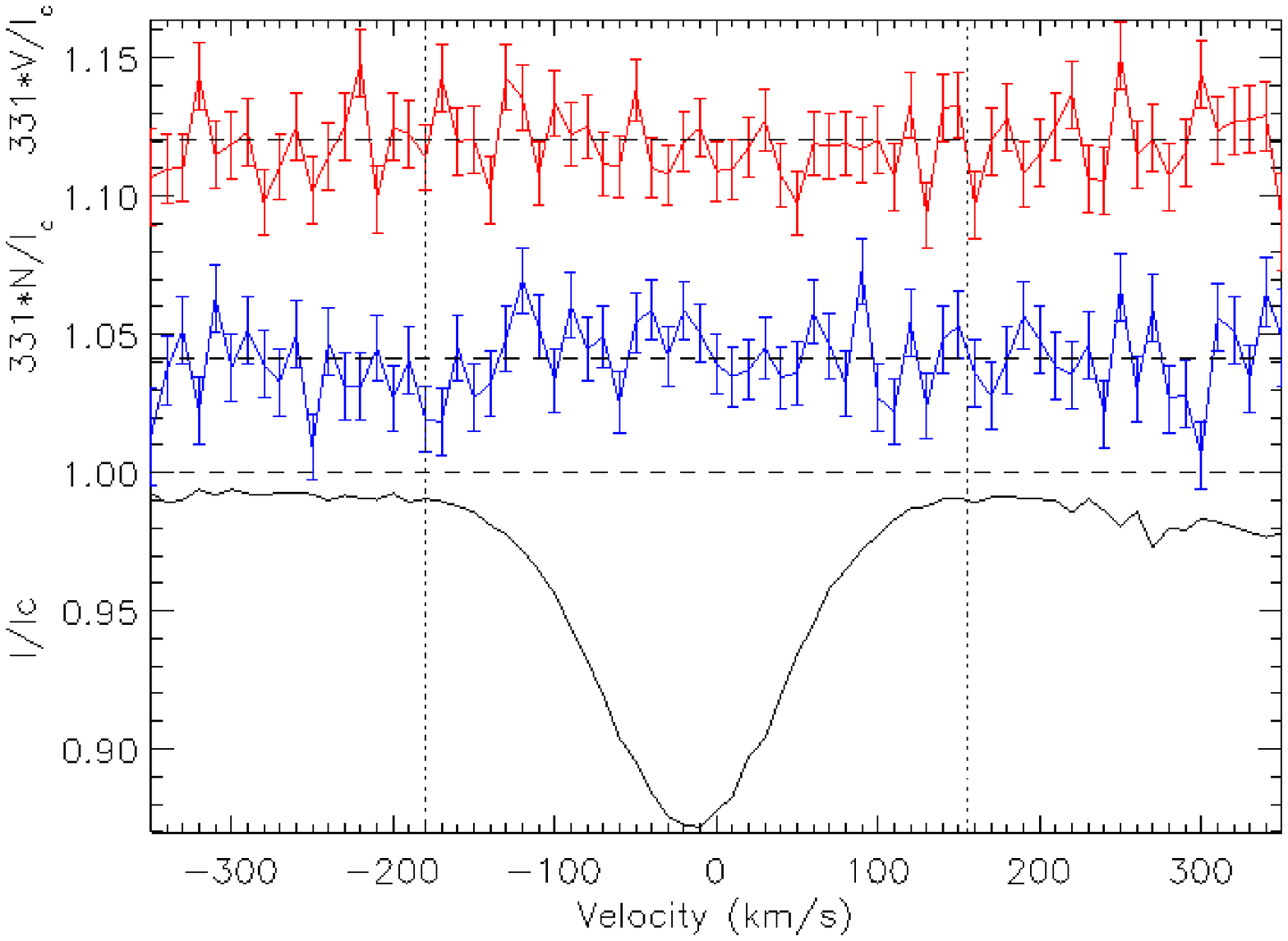}}
\subfigure[$\lambda$ Cep LSD profile.]{\includegraphics[width=2.2in]{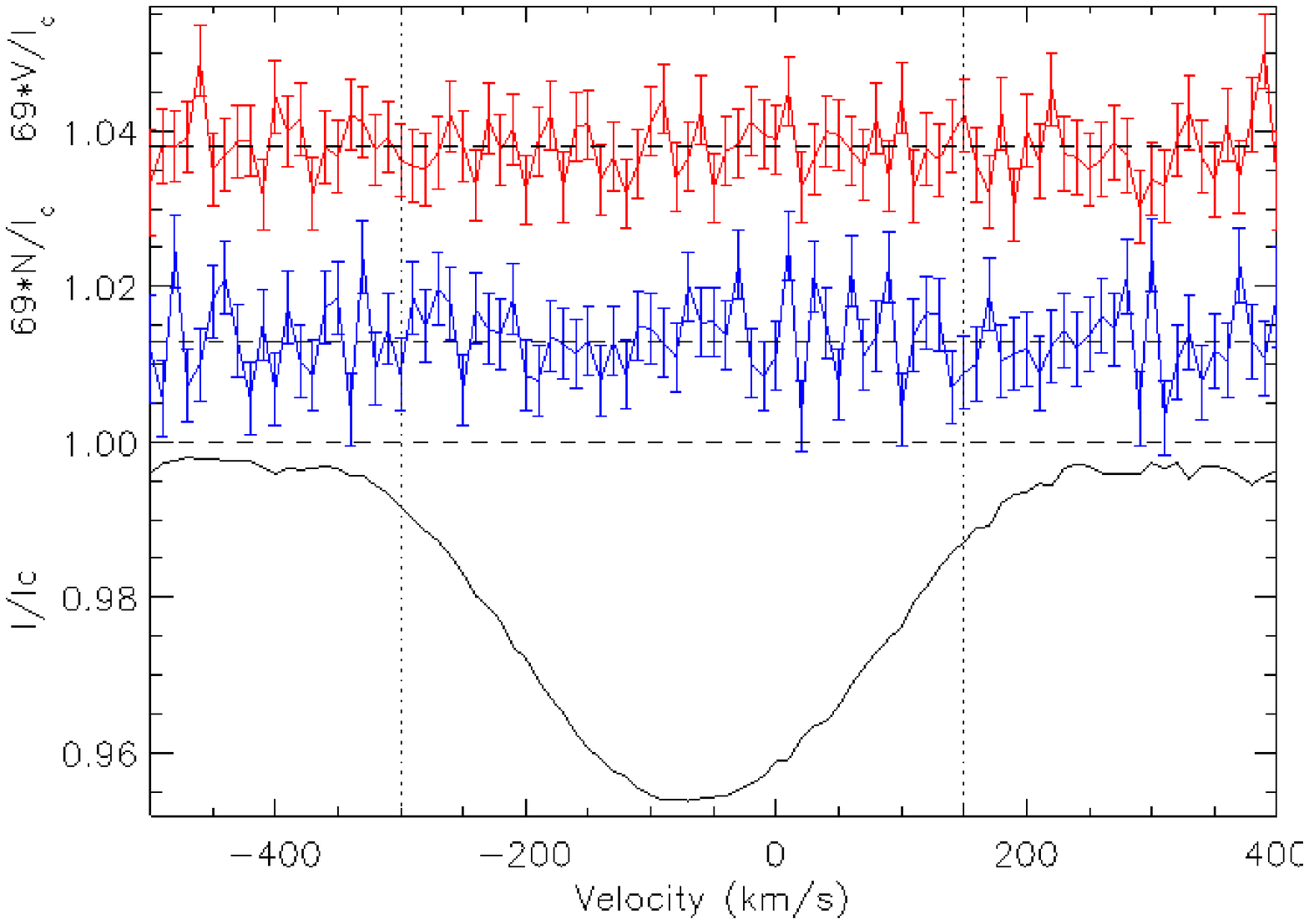}}
\subfigure[10 Lac LSD profile.]{\includegraphics[width=2.2in]{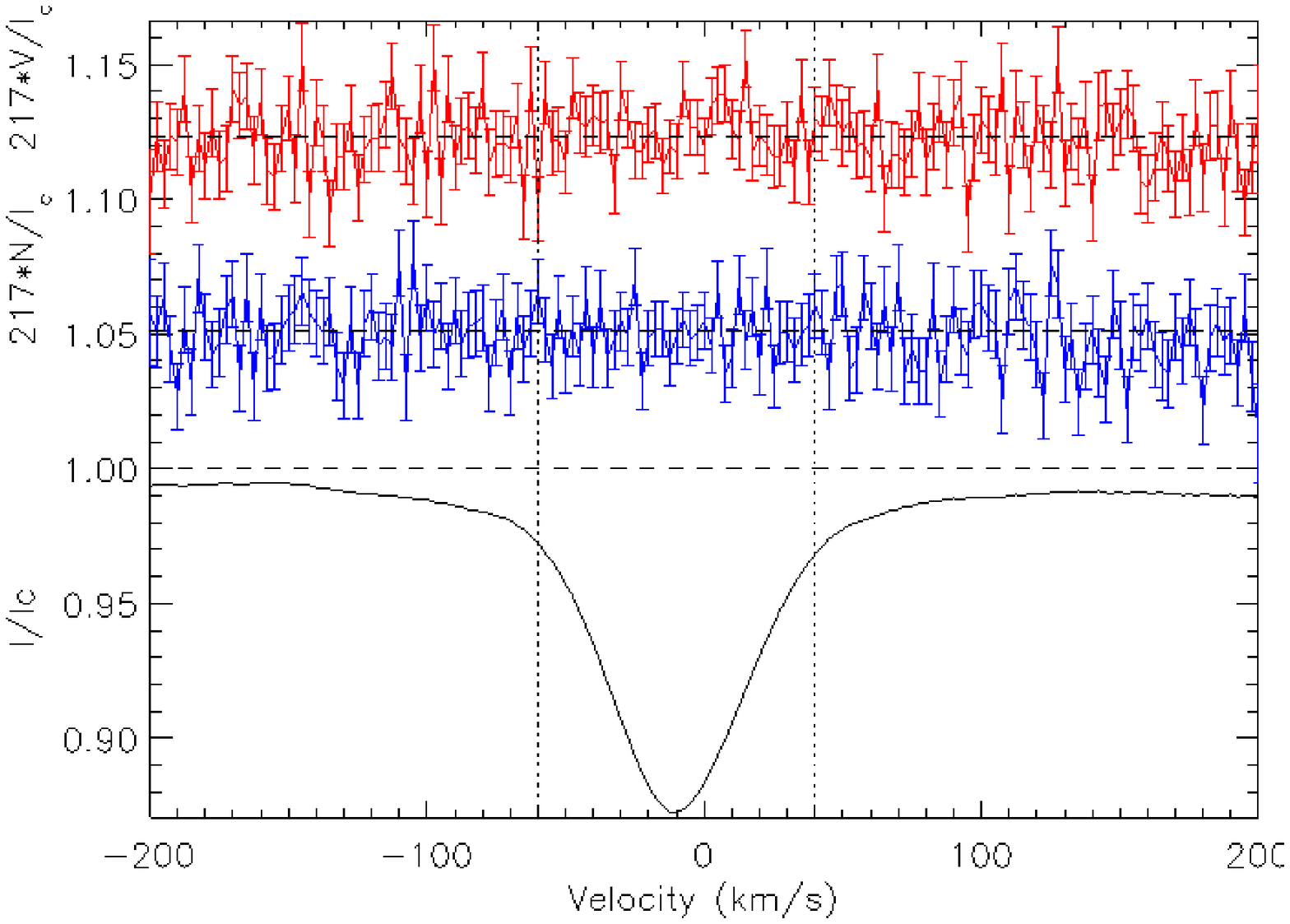}}
\caption{Typical LSD profiles for all stars in the sample. In each plot, the red line (top) is the Stokes V profile, 
while the blue line (middle) is a diagnostic null. Finally, the
black line (bottom) is the Stokes I profile. The dotted lines represent the integration range for each star. 
We can see that no perceptible signal is found in any of the V profiles.}
\label{fig:lsdm}
\end{center}
\end{figure*}

Another measure taken to improve the signal was to co-add the LSD profiles of spectra of each star taken on the same night. The time intervals between the first
and last exposure of a given star on a given night are systematically less than 10\% of the inferred stellar rotational period, therefore
there was no serious risk of smearing the signal and weakening it (see Table~\ref{tab:sample}). A mosaic of sample nightly-averaged LSD profiles for each 
of the stars is presented in Fig.~\ref{fig:lsdm}.

\section{Magnetic field diagnosis}\label{sec:diag}

The LSD profiles were used to assess magnetic fields via two techniques: direct measurement diagnostics and Bayesian inference-based
modeling.

\subsection{Direct measurement diagnostics}

Using the nightly-averaged profiles, as well as the individual ones, the disc-averaged longitudinal magnetic field ($B_{z}$) was computed using
the first-order moments method (e.g. \citealt{b42}). The integration ranges were chosen carefully, after a few trial calculations to determine how to minimize
the error bars without losing any potential signal. Visually, the limits correspond loosely to the zero-points of the second derivative of the Stokes I profiles.
Nightly longitudinal field measurements are listed in Table~\ref{tab:app}. There are no significant detections. Not only do they seem normally
distributed within the error bars, but these error bars are quite small in some cases and provided very tight constraints (e.g. 4 G error bar for 10 Lac
on 17 October 2007). Furthermore, the longitudinal fields are also measured using the diagnostic nulls as a sanity check. On any given night, the error
bars for the longitudinal fields measured from the V profile are consistent with those measured from the nulls, and the distributions of $B_{z}/\sigma_{B_{z}}$
obtained from each profile are essentially identical, which suggests that the V profile does not contain any more signal than the diagnostic nulls.

$\chi^2$ diagnostics are also performed by comparing both the V profile and the diagnostic null
to the null hypothesis ($B = 0$, therefore $V = 0$ and $N = 0$), and detection probabilities are derived from these values \citep{b17}.  These
calculations are performed both within the LSD profile, as well as in the continuum. A detection probability below 99.9\% is considered
as a non-detection, a marginal detection possesses a detection probability between 99.9\% and 99.999\% and a definite detection has a detection probability
of over 99.999\%
(for a discussion of these thresholds, see \citealt{b17}). The 400+ individual and nightly-averaged V profiles are all
non-detections, except 5 cases within the profile (1 in $\zeta$ Oph, 3 in 19 Cep and
1 in 10 Lac) and 1 in the continuum (in $\xi$ Per) all 6 of which are marginal
detections. 
For the ones inside the line, except for a nightly-averaged observation in 10 Lac, the
other 4 occurrences appear in individual observations, with a lower SNR. This could be due to somewhat noisier profiles, and since they are relatively
isolated cases (for all 3 stars there are many more observations which are all non-detections), they are not perceived as being significant. As for the
continuum marginal detection, it is also from a single observation and could be due to noise, as well as slight telluric contamination. On the whole,
these results are largely consistent with those for the diagnostic nulls, further suggesting that there are no real detections.

In summary, both of these direct measurement diagnostics lead to the same conclusion, i.e. that no magnetic field is observed in any of these stars.

\subsection{Bayesian inference}

Additionally, to increase the SNR it is also possible to take advantage of the time resolution provided by repeated measurements. Indeed,
taking into account the oblique dipole rotator model \citep{b19}, data taken at different times should allow to view the surface magnetic field from different
perspectives, thus lifting
some of the degeneracy associated with the geometric parameters of the magnetic field, should it exist. Therefore, using the technique developed by
\citet{b20}, a fully self-consistent Bayesian inference method compares the observed profiles in the Stokes V and N parameters to synthetic Zeeman profiles
for a grid of field strength and geometry parameters. The rotational phase of the observations is also allowed to vary freely, since rotational periods are unknown.

In order to produce synthetic Zeeman profiles to be used for this Bayesian technique, 
it is necessary to estimate the value of the projected rotational
velocity of each star, as well as its macroturbulent velocity. These values are sometimes degenerate and difficult to determine with great precision.
Instead of using previously published values, new values of $v \sin i$ were measured for all stars using the Fourier transform method 
(e.g. \citealt{b30,z3}).
To this effect, synthetic spectra were computed with \textsc{synth}3 \citep{b43}, and the O\textsc{ii} $\lambda 4367$, O\textsc{iii} $\lambda 5508$,
O\textsc{iii} $\lambda 5592$ and C\textsc{iv} $\lambda 5801$ lines were used to compare them to the data. In most cases (10/13), we get relatively 
(e.g. 20\%) lower values of projected
rotational velocity than those reported in the literature \citep{b3}, while for the 4 remaining stars, we get comparable or slightly higher results.
This can be expected, since the line broadening is no longer solely attributed to rotation with this method.

Once the value of $v \sin i$ was determined, the LSD profiles (rather than individual lines, since these are the data we are looking to model)
were then compared to synthetic Voigt profiles to refine the value of $v \sin i$ and determine $v_{mac}$. 
Because this process involved some level of degeneracy, the uncertainty on the obtained values could not be determined in a systematic way, 
but it is conservatively estimated to be about 10-20\%. While this may seem large,
tests using different pairs of values ($v \sin i$ and the associated $v_{mac}$) 
indicate that such a precision is quite sufficient, as errors of this magnitude do not significantly affect the
results of the Bayesian analysis. A summary of these velocity measurements
is given in Table~\ref{tab:bp}, which also contains other relevant physical parameters. 
The macroturbulent velocities are likely to be systematically overestimated; the extra broadening from the helium lines behaves
in a way similar to macroturbulence. However once again, extensive testing on our data has shown that this 
overestimation does not significantly alter the results of the Bayesian inference.

\begin{figure}
\begin{center}
\subfigure[$B_{d}$ PDF for 10 Lac.]{\includegraphics[width=3.4in]{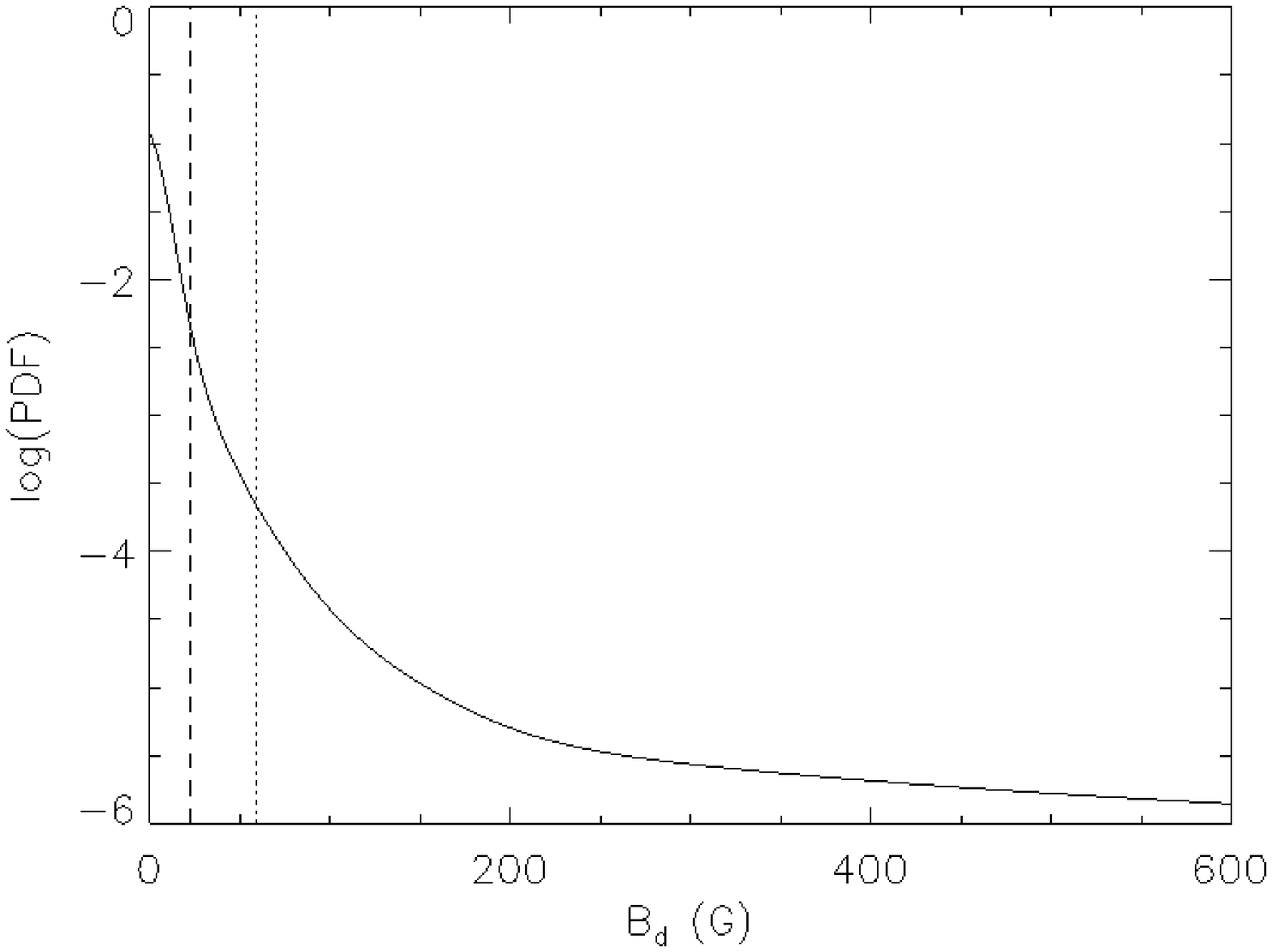}}
\subfigure[$B_{d}$ PDF for $\alpha$ Cam.]{\includegraphics[width=3.4in]{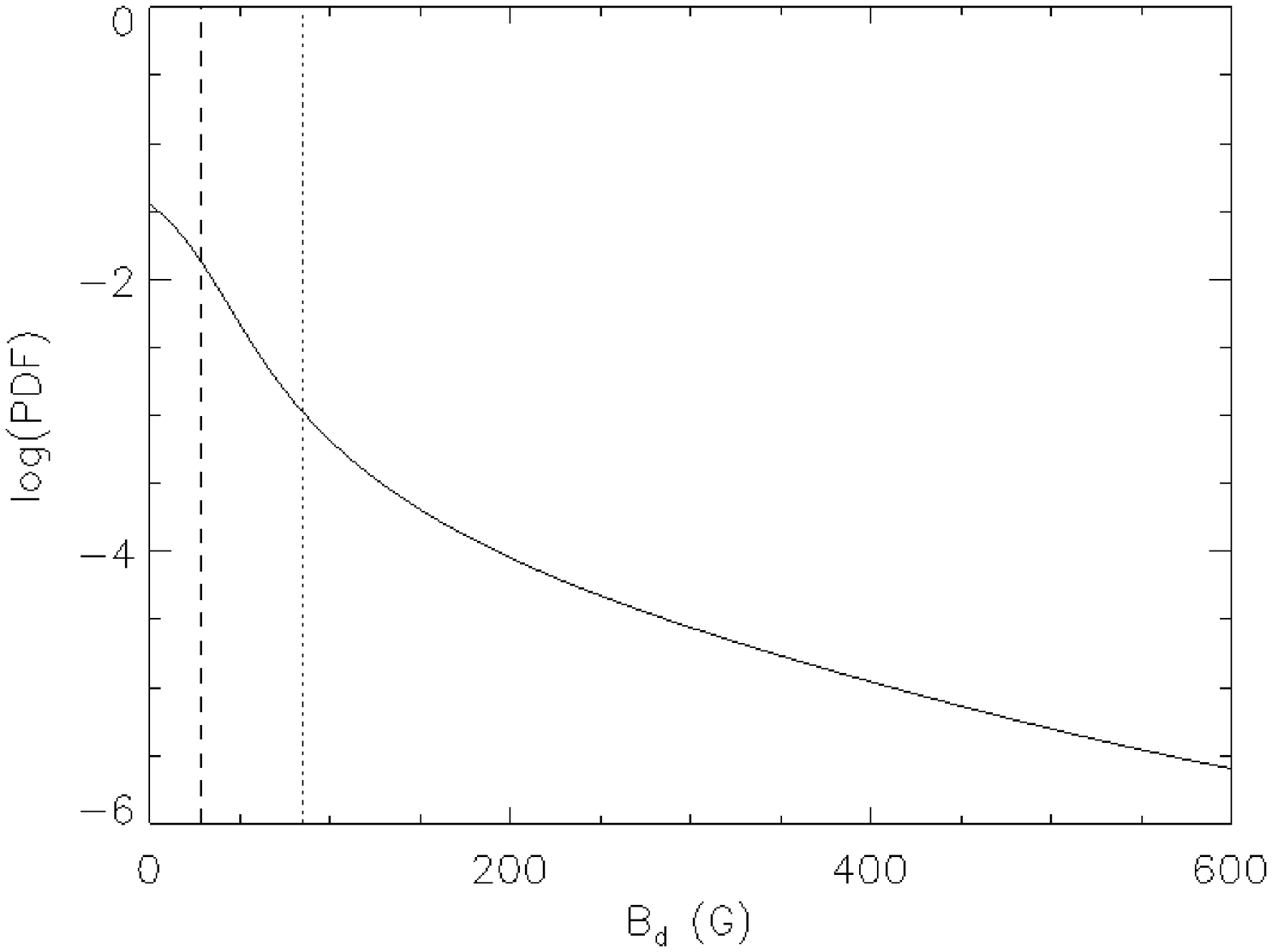}}
\subfigure[$B_{d}$ PDF for HD~34656.]{\includegraphics[width=3.4in]{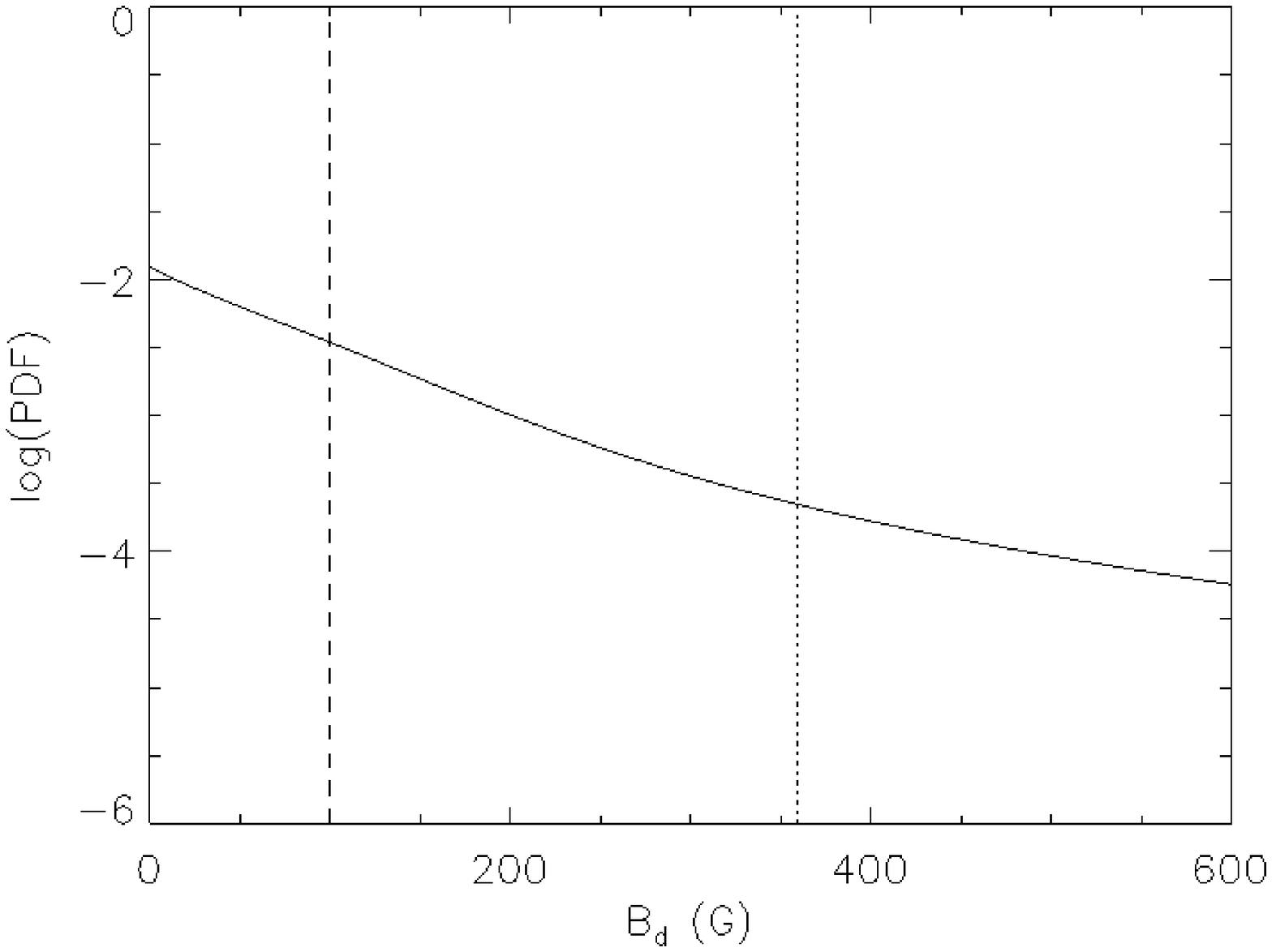}}
\caption{Logarithm of the probability density functions of the dipolar field strength ($B_{d}$) for three representative stars (10 Lac with the best constraints
at the top, $\alpha$ Cam with typical constraints in the middle, and HD34656 with the worst constraints at the bottom)
as derived from the Bayesian inference technique.
For each plot, the dashed line delimits the 68.3\% confidence interval, 
while the dotted line delimits the 95.4\% confidence interval.}
\label{fig:pdfs}
\end{center}
\end{figure}

\begin{figure}
\begin{center}
\includegraphics[width=3.2in]{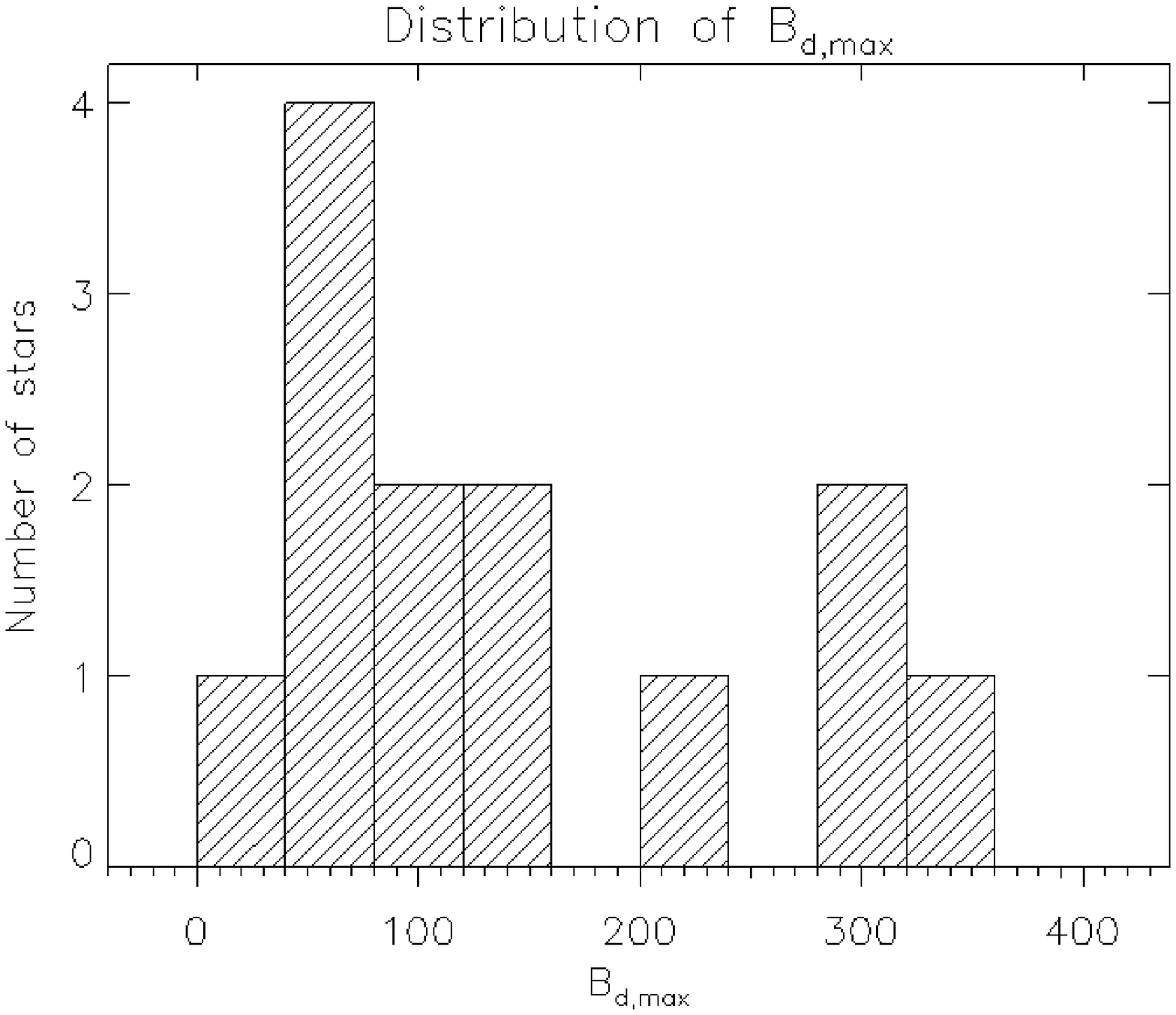}
\caption{Histogram of the $B_{d, \textrm{max}}$ (95.4\% interval upper limit) values derived from the Bayesian analysis
(Table~\ref{tab:bp}). Most stars have an upper limit below 120 G.}
\label{fig:hist}
\end{center}
\end{figure}

Ultimately, we modeled the observed I, V and N profiles to obtain probability 
density functions (PDFs) for 3 variables: the dipolar field strength ($B_{d}$), the inclination angle of the rotational
axis ($i$) and the obliquity angle between the magnetic field axis and the rotational axis ($\beta$). It is also possible to marginalize the PDFs for each variable individually. However, it should be noted that the latter two geometric parameters cannot be constrained in the case
of non-detections \citep{b20}. Figure~\ref{fig:pdfs} shows the marginalized PDFs for three representative stars as a function of $B_{d}$. 
We can see that for each star, the PDF peaks at a value
of 0, which is consistent with a non-detection. Additionally, a similar analysis was performed on the diagnostic null, with consistent results.
Therefore, we obtained no information about the putative field's geometry: we consider the only parameter of interest for this study to be the
strength of the dipolar field. Since we only have non-detections, we can place upper limits on the dipolar field strength by using 
the 95.4\% confidence region upper boundaries (which corresponds to the limit over which we expect the field to be detected, \citealt{b20}). 
These upper limits (noted as $B_{d,\textrm{max}}$) are listed in Table~\ref{tab:bp} (as well as the 68.3\% confidence level upper limits for
comparison purposes). The highest upper limit (95.4\% interval) that we derive is that
of HD 34656 (359 G). This is expected, since there was only a single observation for that star, therefore a lower SNR. The tail of the PDF falls
off less abruptly as well (see Fig.~\ref{fig:pdfs}), since statistically speaking,
the observation could correspond to a particular phase where the field configuration is not suitable for detection. It should be remembered that this technique
aims to take advantage of timeseries of LSD profiles; hence better constraints and a more peaked PDF could be obtained for this star with higher SNR observations
and more extensive time coverage. All the other stars with
upper limits over 100 G (5) have very high projected rotational velocities, which explains their poorer constraints. However, for the rest of the stars (7), we get
extremely tight constraints, in particular in the case of 10 Lac (23 G). These values represent by far the tightest constraints ever obtained for any sample
of OB stars (see Fig.~\ref{fig:hist} for a histogram of these upper limits).

However, even though fast rotating stars have poorer constraints on the strength of their hypothetical dipolar field, their rotation itself suggests that they
do not possess such a field (or if so, a weak one). 
Indeed, a majority of magnetic OB stars are slow rotators. Moreover, all effectively single magnetic O stars are very slow rotators, with periods ranging
from about one week to decades (e.g. \citealt{b22}). This slow rotation is thought to be achieved by the magnetic field, which contributes
to remove angular momentum from a star. This characteristic does not apply to our sample, in which nearly half (6/13) of the stars have projected rotational
velocities of over 200 km/s. We can calculate a typical spindown timescale for a given magnetic field strength (see Eq. 8 of \citealt{asif}).
For example, if we perform that calculation on the supergiant HD~64760 using the 95.4\% interval upper limit on the strength of the field, 
we get a spindown timescale of just under a million years, which seems incompatible with its projected
rotational velocity of 250 km/s.

Another output of the Bayesian analysis is the \textit{odds ratio}. This value represents the ratio of the likelihoods of each of the two hypotheses to be evaluated: 
$H_{0}$, corresponding to no magnetic field, and $H_{1}$, corresponding
to a globally organized dipolar magnetic field. According to \citet{b44}, 
we would need an odds ratio below 1/3 to say that there is weak evidence in favour of the magnetic hypothesis. 
This ratio has been computed for each star (for the individual nightly observations, as well as for the entire
dataset of a given star). 
For all V profiles, we get $odds(H_{0}/H_{1}) > 1$, except for two nightly profiles (1 for $\epsilon$ Ori and 1 for $\xi$ Per), but they do not
go under 0.68. Typical values for the joint datasets range between 1 and 10. None of the stars yield odds ratios favouring the magnetic hypothesis.
These results are also consistent with the odds ratios obtained from the null spectra.

It should be noted that this approach relies on a certain stability of the field. In particular, the geometry and strength of the dipole cannot undergo
significant changes during the period of observation. On the other hand, this method is insensitive to any drift of the dipole in phase (e.g. precession
of the magnetic axis around the rotation axis at a non-uniform rate). We assume that the
geometry of the field remains stable over timescales of at least a few years given the temporal baseline of our observations; this assumption 
is found to be justified in intermediate-mass and massive stars (e.g. \citealt{Wade, Grun, Silv}). In any case, for a majority of stars in the sample,
most observations are grouped within a few months, periods over which secular changes in the field geometry would not be important.

Once again, this analysis supports the view that no magnetic fields are observed, but further allows us to compute quantitative upper limits on the surface
dipole component, necessary for evaluating the potential influence on the stellar wind.

\begin{table*}
\caption[Stellar parameters and results]{Stellar and magnetic parameters of the stars in the sample. Terminal wind 
velocities are obtained from \citet{b3} and references therein, as well as the previously
published values of the projected rotational velocity (in parentheses). New values of $v \sin i$ obtained from the Fourier transform method
(and refined by fitting the profiles) are reported as well.
Nine stars of the sample are studied by \citet{b4} (a) and \citet{b5} (b),
and all their other properties were obtained from these references (in particular, mass-loss rates are obtained using the empirical relation
of \citealt{b6}). 
For the B supergiants ($\epsilon$ Ori and HD~64760), \citet{b10} (c) provide the radii and mass-loss rates, while the remaining
parameters are obtained from \citet{b11} (d). Finally, \citet{b6} (e) provide the radii, mass-loss rates and effective
temperatures of $\zeta$ Pup and $\zeta$ Oph; \citet{b7} (f) detail the DAC recurrence for the former and 
\citet{b8} (g) do the same for the latter. }\label{tab:bp}
\begin{tabular}{|l|c|c|c|c|c|c|c|c|c|c|c|l|}
  \hline
Name & $R_{*}$ & $T_{\mathrm{eff}}$ & $\dot{M}$ & $v_{\infty}$ & $v \sin i$ & $v_{mac}$ & $P_{\mathrm{max}}$ & $t_{\mathrm{DAC}}$ & $B_{d,\textrm{max}}$ & 
$B_{d,68.3\%}$ & $\eta_{*,\textrm{max}}$ & Ref.\\
   & ($R_{\odot}$) & (kK)          & ($M_{\odot}$/yr) & (km/s) & (km/s)  & (km/s) & (d)                  & (d)               & (G) & (G) &  & \\
  \hline
$\xi$ Per       & 11 & 36.0 & $3 \cdot 10^{-7}$  & 2330 & 215~(213)& 80 & 2.6  & 2.0      & 59  & 22  &  0.11 & a \\
$\alpha$ Cam    & 22 & 29.9 & $9 \cdot 10^{-7}$  & 1590 &  90~(129)& 85 & 12.4 & a few    & 85  & 28  &  0.48 & a, b \\
HD~34656        & 10 & 36.8 & $2 \cdot 10^{-7}$  & 2155 &  70~(91) & 65 & 7.2  & 0.9      & 359 & 100 &  5.75 & a \\
$\lambda$ Ori A & 12 & 35.0 & $3 \cdot 10^{-7}$  & 2175 &  55~(74) & 60 & 11.0 & $>$ 5    & 65  & 22  &  0.18 & a \\
$\epsilon$ Ori  & 32 & 28.6 & $2 \cdot 10^{-6}$  & 1910 &  65~(91) & 55 & 24.9 & 0.7      & 78  & 29  &  0.31 & c, d \\
15 Mon          & 10 & 41.0 & $4 \cdot 10^{-7}$  & 2110 &  50~(67) & 53 & 10.1 & $>$ 4.5  & 84  & 30  &  0.16 & a \\
HD~64760        & 23 & 23.1 & $1 \cdot 10^{-6}$  & 1500 & 250~(216)& 50 & 4.7  & a few    & 282 & 89  &  5.37 & c, d \\
$\zeta$ Pup     & 16 & 42.4 & $1 \cdot 10^{-6}$  & 2485 & 220~(219)& 80 & 3.7  & 0.8      & 121 & 34  &  0.29 & e, f \\
$\zeta$ Oph     & 8  & 35.9 & $9 \cdot 10^{-8}$  & 1505 & 375~(372)& 50 & 1.1  & 0.8      & 224 & 75  &  4.57 & e, g \\
68 Cyg          & 14 & 36.0 & $7 \cdot 10^{-7}$  & 2340 & 290~(305)& 65 & 2.4  & 1.3      & 286 & 90  &  1.86 & a \\
19 Cep          & 18 & 30.2 & $6 \cdot 10^{-7}$  & 2010 &  56~(95) & 70 & 16.3 & $\sim$ 5 & 75  & 28  &  0.30 & a \\
$\lambda$ Cep   & 17 & 42.0 & $3 \cdot 10^{-6}$  & 2300 & 200~(219)& 80 & 4.3  & 1.4      & 136 & 50  &  0.15 & a \\
10 Lac          & 9  & 38.0 & $1 \cdot 10^{-7}$  & 1140 &  21~(35) & 30 & 21.7 & $>$ 5    & 23  & 8   &  0.07 & a \\
  \hline
\end{tabular}
\end{table*}

\section{Notes on individual stars}\label{sec:notes}

The following subsections contain notes about each individual star.

\begin{figure}
\begin{center}
\includegraphics[width=3.4in]{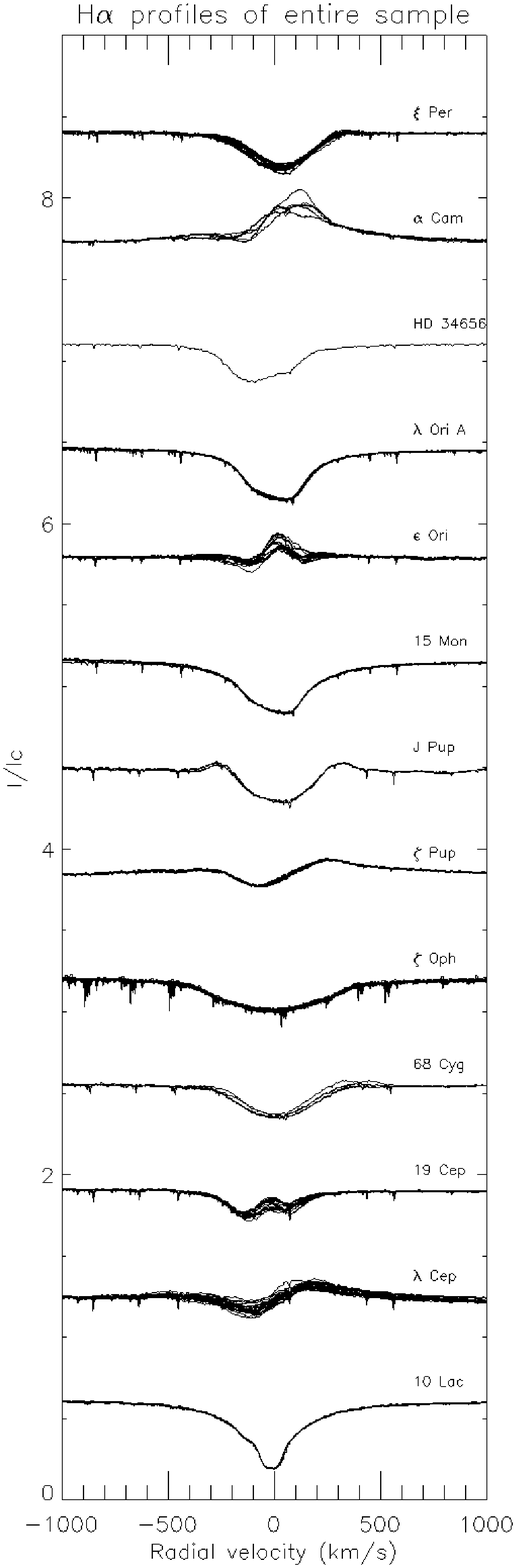}
\caption{H$\alpha$ profiles of all stars (offset for viewing purposes). 
Some stars have very little to no variability, whereas others have significant variability (variable depths, emission, etc.).}
\label{fig:halpha}
\end{center}
\end{figure}

\subsection{$\xi$ Per}

$\xi$ Per is a well-known O7.5 giant runaway star \citep{x1} whose DAC behaviour has been extensively studied in the past (e.g. \citealt{b4}). 
\citet{b31} have studied its spectral
variability in a number of wind-sensitive lines and also confirm the presence of NRPs. While its high projected rotational velocity makes it
harder to perform precise magnetometry, the excellent time coverage of this dataset leads to a very tight upper limit on the strength of
an hypothetical dipolar field. There does not seem to be significant variation in the shape of H$\alpha$ during our observing runs, but rather simply a modulation of
the depth of the line (see Fig.~\ref{fig:halpha} for a summary of the H$\alpha$ profiles of all stars). 
Forty-four independent observations of $\xi$ Per were acquired over 13 nights in December 2006, September 2007
and November 2011. The smallest nightly longitudinal field error bar calculated from these data is 21 G, and the derived dipolar field strength upper limit is
59 G.

\subsection{$\alpha$ Cam}

Also a runaway \citep{x1}, $\alpha$ Cam (O9.5 supergiant) exhibits a subtler DAC behaviour \citep{b5}. The projected rotational velocity
was significantly revised (see Table~\ref{tab:bp}). 
The H$\alpha$ profile undergoes important changes from night to night. Eleven independent observations of $\alpha$ Cam were acquired over 5 nights 
between 2006 and 2013. The smallest nightly longitudinal field error bar calculated from these data is 10 G, and the derived dipolar field strength upper 
limit is 85 G.

\subsection{HD~34656}

HD~34656 is a well-studied O7 bright giant (e.g. \citealt{x2}, who observed line profile variations in its spectra)
with relatively low $v \sin i$, making it an interesting target for this kind of study. \citet{b4} have characterized its DAC behaviour.
Unfortunately, there
was only a single observation of the star in the archive, therefore it was not possible to constrain its magnetic properties with great precision.
The observation of HD~34656 was acquired in November 2011.
The longitudinal field error bar calculated from this observation is 38 G, and the derived dipolar field strength upper limit is
359 G.

\subsection{$\lambda$ Ori A}

In a large separation double system with an early-B star (e.g. \citealt{x3}), 
$\lambda$ Ori A is a slowly-rotating O8 giant, exhibiting well-known DAC behaviour (e.g. \citealt{b4}). We placed a very
firm upper limit on its dipolar field strength. No detectable variations are found in H$\alpha$ in our observations.
Twenty independent observations of $\lambda$ Ori A were acquired over 8 nights between 2007 and 2010.
The smallest nightly longitudinal field error bar calculated from these data is 12 G, and the derived dipolar field strength upper limit is
65 G.

\subsection{$\epsilon$ Ori}

One of two B supergiants present in this sample, the DAC behaviour of $\epsilon$ Ori (B0) 
was first described by \citet{b11}. Evidence suggesting the possible presence
of NRPs is offered by \citet{b34}. We derive rather tight magnetic constraints, on top of observing significant variations of the H$\alpha$ profile over time.
Seventy independent observations of $\epsilon$ Ori were acquired over 9 nights in October 2007, October 2008
and March 2009. The smallest nightly longitudinal field error bar calculated from these data is 6 G, and the derived dipolar field strength upper limit is
78 G.

\subsection{15 Mon}

A long period spectroscopic binary \citep{b32} with well-studied DACs \citep{b4}, 
15 Mon (O7 dwarf) has low $v \sin i$, thus leading to a well-constrained field upper limit, even though
it has not been observed as extensively as some other stars in this sample. 
Our observations of 15 Mon do not present noticeable changes in H$\alpha$.
Contrarily to \citet{b37}, who claimed a 4.4 $\sigma$ detection based on two
observations with FORS2 and SOFIN (longitudinal field error bars of 37-52 G), we do not find evidence supporting
the presence of a large-scale dipolar magnetic field despite better quality data and more numerous observations. 
Indeed, sixteen independent observations of 15 Mon were acquired over 8 nights in December 2006, September-October 2007
and February 2012. The smallest nightly longitudinal field error bar calculated from these data is 20 G, and the derived dipolar field strength upper limit is
84 G.

\subsection{HD~64760}

This B0.5 supergiant was studied by \citet{b14}, who not only detect DACs, but also other forms of variability such as ``phase bowing", making this star
a complex but very interesting case. It is also known to exhibit signs of NRPs (e.g. \citealt{b33}). However, due to its high projected rotational
velocity, as well as the low number of observations, its magnetic properties are amongst the worst-constrained of this sample. There is no variation of H$\alpha$
between the two nights it was observed.
Nine independent observations of HD~64760 were acquired over 2 nights in November 2010
and December 2012. The smallest nightly longitudinal field error bar calculated from these data is 37 G, and the derived dipolar field strength upper limit is
282 G.

\subsection{$\zeta$ Pup}

Characterized by a very strong wind, $\zeta$ Pup is a particularly hot O4 supergiant. Its DAC behaviour was evidenced by \citet{b7}, while \citet{b36} suggest
the possibility of NRPs. We provide good limits on the magnetic field, albeit with a single night of observations. Better time coverage could provide much
better constraints. It is not obvious from these data whether the H$\alpha$ profile varies over the course of the night.
Thirty independent observations of $\zeta$ Pup were acquired over a single night in February 2012.
The nightly longitudinal field error bar calculated from these data is 21 G, and the derived dipolar field strength upper limit is
121 G.

\subsection{$\zeta$ Oph}

A well-known runaway star (e.g. \citealt{x4}), $\zeta$ Oph (O9.5 dwarf) possesses a very high value of $v \sin i$ and short-period DACs \citep{b8}. 
Nonetheless, thanks to great time coverage, we obtain good magnetic constraints.
\citet{b37} claim this star to be magnetic, a result we do not reproduce here. Although their nightly observations possess better individual error
bars, their longitudinal field curve has an amplitude of roughly 120 G and implies a surface dipole field of at least 360 G, 
which seems inconsistent with the 224 G upper limit
we place on $B_{d}$. Period analysis performed on our longitudinal field measurements (for V and N) 
with \textsc{period04} \citep{j1} does not suggest periodic behaviour; in particular,
the 0.8d and 1.3d periods reported by \citet{b37} are not recovered. The periodogram of both the Stokes V and the null results are quite similar, further suggesting
that no periodic signal is to be found. Individual Stokes I LSD profiles show strong
line profile variations (LPV), in the form of bumps appearing and disappearing across the profile, which are indicative of the presence of NRPs,
known to exist in this star (e.g. \citealt{z4}). 
We do not detect noticeable variations in H$\alpha$ from night to night in our runs.
Sixty-five independent observations of $\zeta$ Oph were acquired over 46 nights in 2011 and 2012.
The smallest nightly longitudinal field error bar calculated from these data is 118 G, and the derived dipolar field strength upper limit is
224 G.

\subsection{68 Cyg}

The O7.5 runaway (e.g. \citealt{j2}) giant 68 Cyg is a rapid rotator with well-studied DACs \citep{b4}. 
Factoring that in with a small number of observations, the putative dipolar magnetic field strength of 
68 Cyg is not as well constrained as most of the other stars of the sample. However, H$\alpha$ is seen to be variable, though the pattern of its
variation with time is not clear.
Eight independent observations of 68 Cyg were acquired over 4 nights between 2006 and 2012.
The smallest nightly longitudinal field error bar calculated from these data is 46 G, and the derived dipolar field strength upper limit is
286 G.

\subsection{19 Cep}

Believed to be a multiple star system \citep{b39}, 19 Cep is known to exhibit DAC behaviour \citep{b4} and has a primary (O9.5 supergiant) 
with low projected rotational velocity, so it was possible to obtain a firm upper limit on the dipolar
magnetic field. The H$\alpha$ profiles show some signs of variability.
Thirty-three independent observations of 19 Cep were acquired over 10 nights between 2006 and 2010.
The smallest nightly longitudinal field error bar calculated from these data is 17 G, and the derived dipolar field strength upper limit is
75 G.

\subsection{$\lambda$ Cep}

The hot (O6) supergiant $\lambda$ Cep is a runaway (e.g. \citealt{j2}) 
with a high value of $v \sin i$ and relatively short-period DACs \citep{b4}. Extensive time coverage leads to good magnetic constraints, despite
the fast rotation.
This star is also believed to harbour NRPs (e.g. \citealt{b40}). Strong variations of the H$\alpha$ profile are observed.
Twenty-six independent observations of $\lambda$ Cep were acquired over 26 nights between 2006 and 2012.
The smallest nightly longitudinal field error bar calculated from these data is 57 G, and the derived dipolar field strength upper limit is
136 G.

\subsection{10 Lac}

Hosting weaker (but detectable) wind variations \citep{b4}, 10 Lac is a sharp-lined O9 dwarf, leading to exceptionally tight limits on the field strength. No
H$\alpha$ variations are detected in our data. 
Thirty-six independent observations of 10 Lac were acquired over 18 nights in December 2006, September-October-November 2007
and July 2008. The smallest nightly longitudinal field error bar calculated from these data is 4 G, and the derived dipolar field strength upper limit is
23 G, both of which are the best constraints obtained for any star in this sample.

\section{Discussion and Conclusions}\label{sec:disc}

As shown in the previous sections, no large-scale dipolar magnetic field is detected in any of the 13 stars of this sample. However,
in order to draw conclusions on whether such fields could be the cause for DACs, it is important to investigate the different possible
interactions between weak, potentially undetected magnetic fields and stellar winds. 

One form of interaction that has been increasingly investigated in the past years is magnetic wind confinement. Indeed, the magnetic field
can channel the wind and closed loops can effectively ``confine'' it, leading to material trapped in a magnetosphere of closed magnetic loops. \citet{b21} 
introduce the following ``wind confinement" parameter to characterize this interaction:

\begin{equation}
\eta_{*} = \frac{{B_{eq}}^2 {R_{*}}^2}{\dot{M} v_{\infty}}
\end{equation}

\noindent where $B_{eq}$ corresponds to the strength of the magnetic field at the equator (which equals half of the dipole polar field strength, $B_{d}$),
$R_{*}$ is the stellar radius, $\dot{M}$ is the mass-loss rate and $v_{\infty}$ is the terminal velocity of the wind. In effect, this parameter
corresponds to the ratio of the magnetic field energy density and the wind kinetic energy density at the stellar surface; therefore, its value gives a sense of
which of the two dominates. If $\eta_{*} << 1$, then the wind's momentum causes the magnetic field lines to stretch out radially and the outflow is essentially
unperturbed. On the other hand, if $\eta_{*} >> 1$, then the strong magnetic field lines are perpendicular to the outflow at the star's magnetic equator, barring
the passage of charged material. Depending on the rotational parameters of the star, this can lead either to a centrifugal or a dynamical magnetosphere
(for a detailed description of both these cases, see \citealt{b22}).

In intermediate cases however, the effect of the magnetic field can be somewhat more subtle. An in-depth analysis of this regime is presented
by \citet{b21} and leads to two main thresholds:

\begin{itemize}
\item for $\eta_{*} > 1$, the wind is considered to be confined by the magnetic field;

\item for $0.1 < \eta_{*} < 1$, the wind is not confined, but its flow is significantly affected by the magnetic field.
\end{itemize}

\noindent Therefore, we will consider that for $\eta_{*} < 0.1$, the dynamical effect of the magnetic field on the wind is likely to be too weak to cause DACs.
An upper limit on the value of the $\eta_{*}$ parameter was computed for each star of the sample ($\eta_{*, \textrm{max}}$)
using the upper limit on $B_{d}$ derived from
the Bayesian inference, and the results are presented in Table~\ref{tab:bp}.

It should be noted here that the wind parameter values used to compute these $\eta_{*}$ upper limits are determined empirically. 
For magnetic stars, it is necessary to use theoretical mass-loss rates instead of observed values to represent the net surface driving force, since a significant
part of the outflow can be confined by the magnetic field, and would then not be detected at larger radii \citep{b22}. However, in the case of apparently non-magnetic
stars, the picture is not so clear. Furthermore, our empirical values are found to be systematically comparable to or smaller than theoretical values;
since we are deriving conservative constraints, it seemed more consistent to use the overall smaller empirical values. Finally, while it might be argued that
there are important uncertainties associated with empirical determinations of wind parameters, 
theoretical prescriptions (such as \citealt{b48}) propagate the rather sizeable uncertainties on masses and luminosities, so there
is no obvious reason to choose one over the other based on such an argument.

\begin{figure*}
\begin{center}
\includegraphics[width=6.6in]{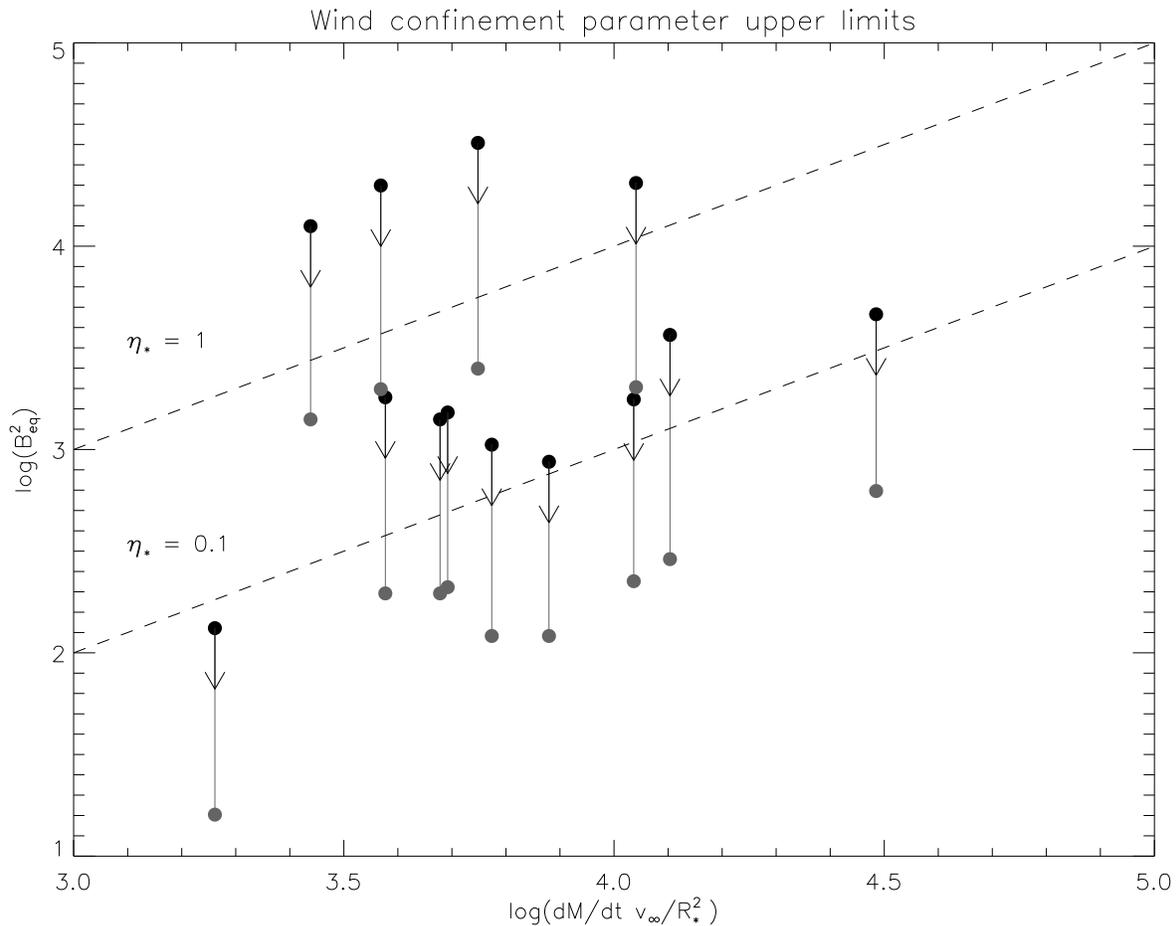}
\caption{Comparison of the magnetic field energy density upper limits (vertical axis, 95.4\% confidence interval upper limits indicated by black points, 
68.3\% confidence interval upper limits indicated by grey points) 
and the wind kinetic energy density values (horizontal axis) for all
13 stars of this study. Dashed lines show where $\eta_{*} = 1$ and
$\eta_{*} = 0.1$. For most stars, the likelihood is greater than 95.4\% that $\eta_{*}$ is below 1, and greater than 68.3\% that it is below 0.1.}
\label{fig:eta}
\end{center}
\end{figure*}

The value of $\eta_{*, \textrm{max}}$
ranges between 0.072 and 5.75, with one star below a value of 0.1 (10 Lac) and a majority of the stars below
a value of 1 (9/13). As for the stars with $\eta_{*, \textrm{max}} > 1$, they all have
very high projected rotational velocities, thus making it difficult to tightly constrain the field strength. These results are also illustrated in
Fig.~\ref{fig:eta}, where the x-axis corresponds loosely to the wind kinetic energy density and the y-axis corresponds to the magnetic energy
density. The dashed lines represent our two chosen thresholds. Given the fact that the represented values all correspond to upper limits, 
we can infer that at least a few of these stars do not have magnetic fields strong enough to dynamically affect the wind outflow on the equatorial
plane (as also evidenced by the 68.3\% confidence interval upper limits).

\begin{figure}
\begin{center}
\subfigure[$\eta_{*}$ cumulative PDF.]{\includegraphics[width=3.4in]{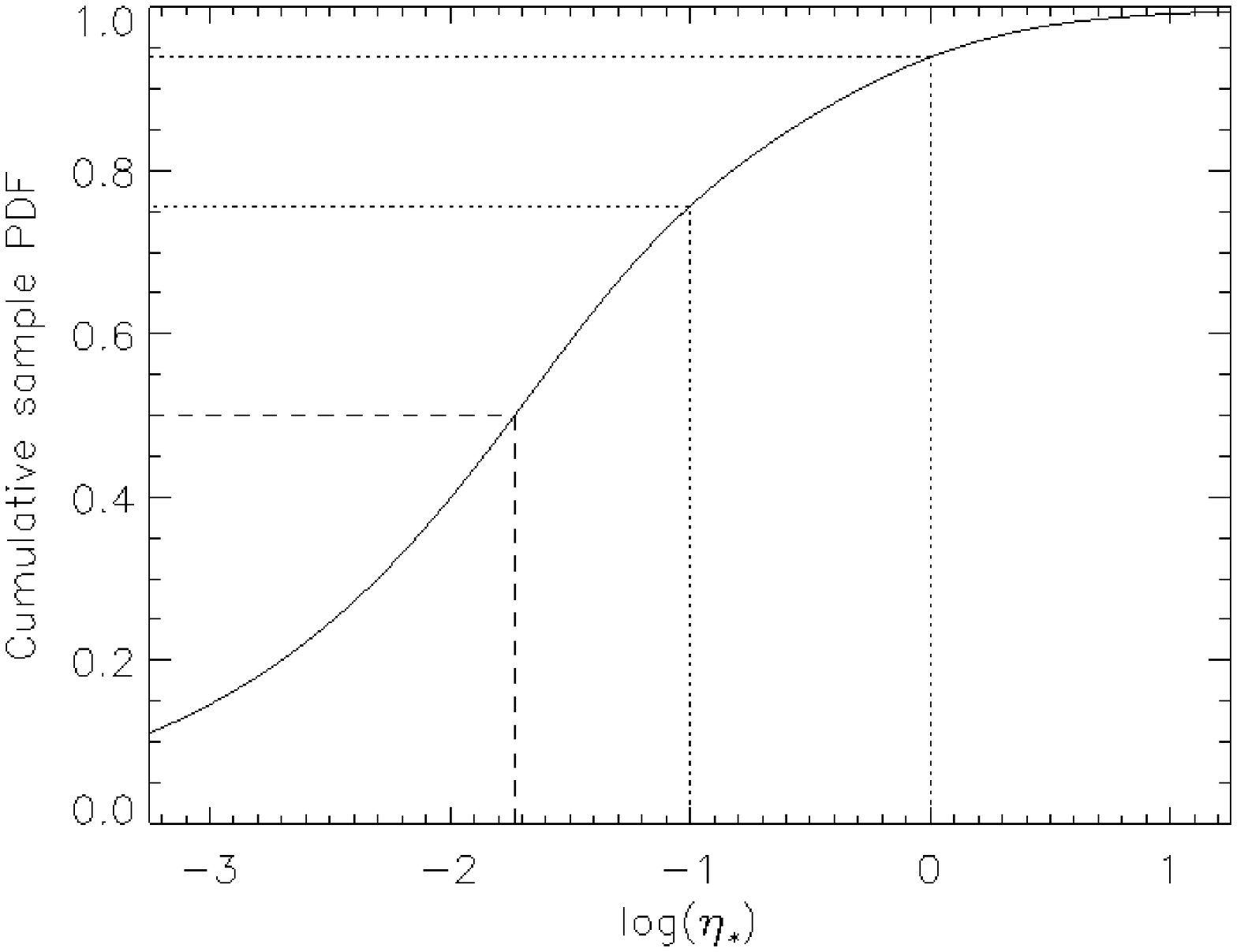}}
\subfigure[$B_{d}$ cumulative PDF.]{\includegraphics[width=3.4in]{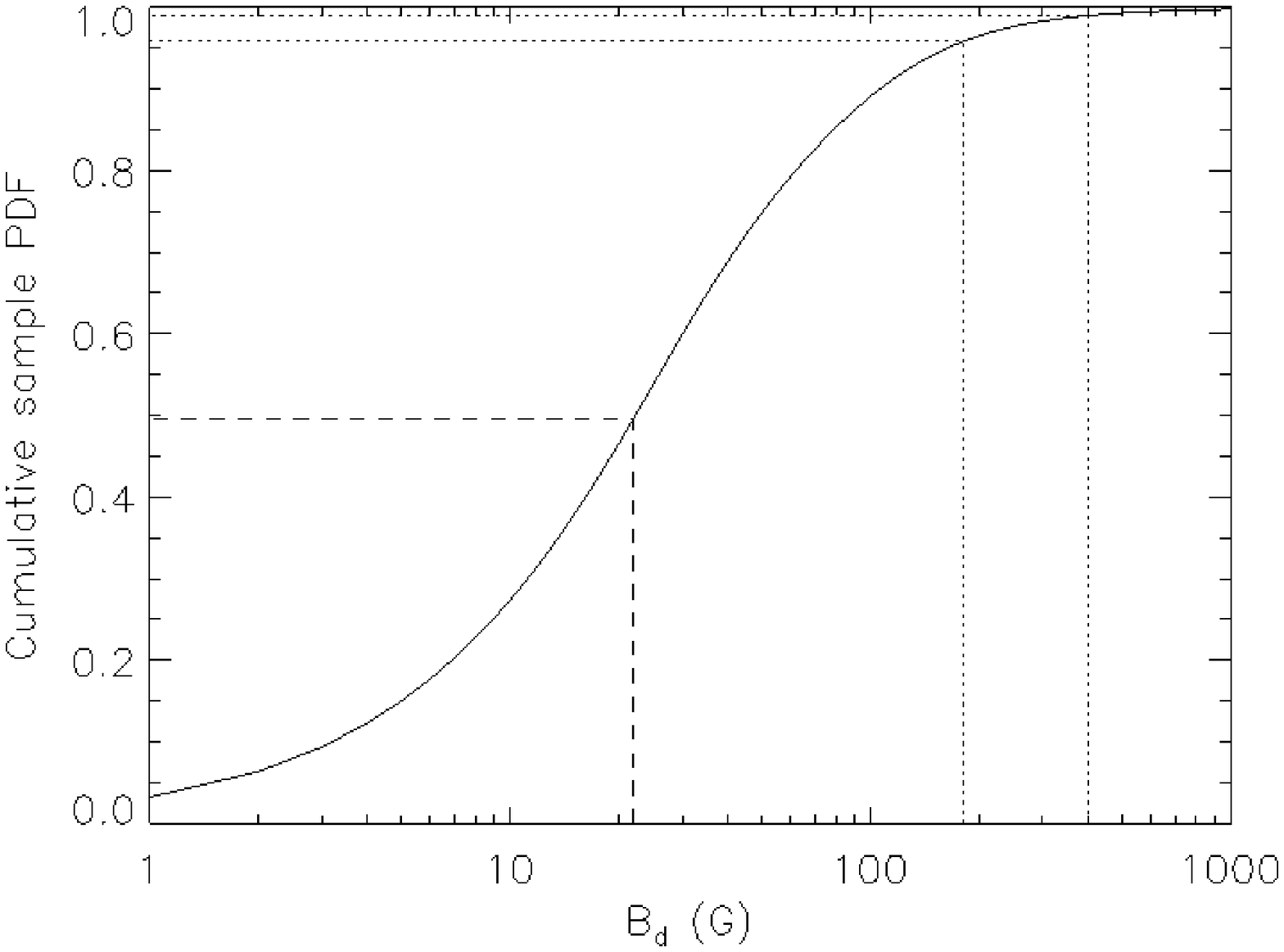}}
\caption{Cumulative PDFs of the total sample for $\eta_{*}$ (top) and $B_{d}$ (bottom). In both cases, the dashed line shows the 50\% confidence
interval upper limit. For the top panel the dotted lines represent, from left to right, $\eta_{*}=0.1$ and $\eta_{*}=1$. For the bottom panel the dotted
lines represent, from left to right, the field strength required to produce a 10\% and a 50\% brightness enhancement (resp. about 180 G and 400 G).}
\label{fig:cum_pdfs}
\end{center}
\end{figure}

In addition to the upper limits, we use the PDFs to assess the sample's distribution of confinement, assuming that each star contributes
probabilistically to various field strength bins according to its normalized probability density function (constructing, in other words,
a ``probabilistic histogram" of field strengths). In this way, we account for both the most probable
field strength as well as the large-field tail of the distributions. The top panel of Figure~\ref{fig:cum_pdfs} shows this
global cumulative PDF for the wind confinement parameter. We expect any star selected from the sample to have $\eta_* < 0.02$ (which is well
below the threshold of $\eta_* = 0.1$) with
a probability of 50\%, or in other words, we expect half of the sample to have a confinement parameter value
below 0.02. Using this
cumulative PDF, we can also calculate that 75.6\% of the sample should
have $\eta_* < 0.1$ and 93.9\% of the sample should have $\eta_* < 1$.
Assuming this small sample
is representative of the larger population of stars displaying DACs, this
implies that there is no significant dipolar magnetic
dynamic influence on the wind for most of these stars.
Under these conditions, wind confinement by a dipolar magnetic field
does not seem to be a viable mechanism to produce DAC-like variations in
all stars.


The derived values of $\eta_{*}$ are sensitive to uncertainties in the values of $R_{*}$, $\dot{M}$ and $v_{\infty}$. While the last parameter
is essentially identical in all studies, in some extreme cases values of $R_{*}$ can be up to 2-2.5 times larger than the adopted values, 
whereas $\dot{M}$ can be up to 10 times larger. Such differences would result respectively in a 6-fold increase and a 
10-fold decrease in the inferred value of $\eta_{*}$. However, studies that infer larger stellar radii also tend to infer larger mass loss rates 
(e.g. \citealt{plouc1}, with $\xi$ Per and HD 34656). 
Thus one effect approximately offsets the other.
The largest potential increase in inferred $\eta_{*}$ for a star of our sample would occur for 
$\alpha$ Cam; based on the values measured by \citet{plouc3} (about 1.5 times increase
in radius, and half the mass-loss rate), we obtain an increase of $\eta_{*}$ by a factor of 4. However, for typical combinations of $R_{*}$ and $\dot{M}$ obtained
from other studies, we obtain values of $\eta_{*}$ that are either comparable in magnitude, or smaller (up to an order of magnitude) 
than those inferred using the adopted parameters.

Do these results rule out dipolar fields altogether? \citet{b16} simply introduce bright spots on the surface of the star, with no particular attention to
the mechanism creating these. While wind confinement is possibly the most obvious effect of a magnetic field on the ouflowing material driven from the
surface of a massive star, there might also be more subtle interactions. For instance, the magnetic pressure at the poles of a weak large-scale dipolar 
field could lower the local gas pressure, thus reducing the gas density and leading to a lower optical depth. Hence, light coming from the pole
would actually probe hotter regions within the star. 
This could cause bright spots like those in the \citet{b16} model. Making a few assumptions (closely modeled on the
calculations of \citealt{b27}), we can derive a simplified
formula for the magnetic field ($B$) required to produce a given brightness enhancement. Indeed, if we consider a flux tube at the photosphere, we can
compare a zone outside of the tube ($B = 0$) to a zone inside the tube ($B = B_{\textrm{T}}$). Furthermore, we assume a grey atmosphere:

\begin{equation}
T(\tau) = T_{\textrm{eff}}\left( \frac{3}{4}\tau + \frac{1}{2} \right)^{\frac{1}{4}}
\end{equation}

\noindent where $T$ is the temperature and $T_{\textrm{eff}}$ is the effective temperature (corresponding to an optical depth, $\tau$,
of 2/3). At equilibrium, the gas pressures ($P_{\textrm{g}}$) inside and outside the tube only differ by the value of the
magnetic pressure ($P_{\textrm{B}} = \frac{B^{2}}{8 \pi}$):

\begin{equation}
P_{\textrm{g}}(r) = P_{\textrm{g}}'(r) + P_{\textrm{B}}
\end{equation}

\noindent where primed variables refer to values inside the flux tube, by opposition to unprimed variables
which refer to values outside the flux tube. The optical depth can be written as a function of gas pressure:

\begin{equation}
\tau = \frac{\kappa P_{\textrm{g}}}{g}
\end{equation}

\noindent where $\kappa$ is the mean Rosseland opacity, and $g$ is the surface gravity. To determine the brightness enhancement, we need to find the 
temperature corresponding to an optical depth of 2/3 inside the flux tube (assuming magneto-hydrostatic and temperature equilibrium at a given vertical
depth):

\begin{equation}
T(\tau' = 2/3) = T_{\textrm{eff}} \left( 1 + \frac{3 \kappa B^{2}}{32 \pi g} \right)^{\frac{1}{4}}
\end{equation}

\noindent Finally, since the flux is proportional to the fourth power of the temperature, the brightness enhancement can be expressed as:

\begin{equation}
\frac{F'}{F} = 1 + \frac{3 \kappa B^{2}}{32 \pi g}
\end{equation}

Now, using typical values for O dwarfs ($\kappa \sim 1$ and $\log g = 4.0$), it is very simple to perform sample calculations. For instance,
the main model used by \citet{b16} uses a 50\% enhancement. The field required to produce such an enhancement is of the order of 400 G,
assuming a magnetic region surrounded by an adjacent non-magnetic region.
On the other hand, the same paper shows that DAC-like behaviour can arise with an enhancement as small as 10\%. The associated field
would be of the order of 180 G\footnote{It is likely, given that dipole fields correspond to a continuous distribution of magnetic field (rather than isolated
flux tubes as assumed in this calculation), that even stronger polar fields would actually be necessary to achieve a given brightness enhancement.}. 
The dipolar field upper limits shown in Table~\ref{tab:bp} are almost all (9/13) under that value. While models with
smaller brightness enhancements are not tested in their study, this mechanism associated with dipolar magnetic fields does not provide a 
viable way of producing DACs given the observational constraints obtained in this study.

Once again, in very much the same way as we did for $\eta_{*}$, we can compile a global cumulative PDF for $B_{d}$ (bottom panel of Figure~\ref{fig:cum_pdfs}).
The results are quite telling: 50\% of the sample should have $B_d <$ 23 G, and 95.8\% (99.0\%) of the sample should have a smaller dipolar field
strength value than that required to produce a 10\% (50\%) local brightness enhancement.

Even if dipolar fields seem to be an unlikely cause for DACs, the general case for magnetism is not settled. Indeed, structured small-scale magnetic
fields could arise as a consequence of the subsurface convection zone caused by the iron opacity bump at $T \simeq 150 \textrm{kK}$ \citep{b24}. 
Then, magnetic spots at the surface
of the star could possibly give rise to CIRs (e.g. \citealt{c1}), even though they are expected to be relatively weak
(to have a surface field of at least 160 G, we need a 40+ $M_{\odot}$ star). While the detection of such fields is an arduous task
\citep{b25}, proving their existence and understanding their structure might hold the key to this old problem, as well as other
similar problems (e.g. in BA supergiants, see \citealt{shultzinator}). Good candidates for follow-up
deep magnetometry might be $\epsilon$ Ori and 10 Lac. The former has the advantage of being very bright and having a relatively low value of $v \sin i$, while the
latter has very low projected rotational velocity (for an O star). 10 Lac already has decent time coverage, but could benefit from obtaining more observations per
night.

In parallel to observational efforts, theoretical investigations
are required in order to probe the parameter space of magnetic field strengths and configurations 
to find out which types of fields can give rise to DAC-like phenomena. Numerical simulations can also be used to investigate mechanisms other than magnetism, 
as well as constrain the required brightness
enhancement in a \citet{b16} model analog to create CIRs in the first place.

The next paper of this series will explore the magnetic spot hypothesis and hopefully place constraints on how likely such a mechanism is to
cause DACs.

\section*{Acknowledgments}

This research has made use of the SIMBAD database operated at CDS, Strasbourg, France and 
NASA's Astrophysics Data System (ADS) Bibliographic Services.

ADU gratefully acknowledges the support of the \textit{Fonds qu\'{e}b\'{e}cois de la recherche sur la nature et les technologies}. GAW is supported by an NSERC
Discovery Grant. AuD acknowledges support from the NASA Chandra theory grant to Penn State Worthington Scranton and NASA ATP Grant NNX11AC40G.

Finally, the authors thank the anonymous referee for his insightful comments which have no doubt 
contributed to making this paper better.

\appendix

\onecolumn
\begin{center}

\section{List of observations}

\begin{longtable}{ c c r r c c }
\caption[List of observations]{Full list of nightly observations for each star. The date is given in universal time (UT), $B_{z}$ is the nightly measured
longitudinal magnetic field value, ${\sigma}_{B_{z}}$ is the nightly error bar on the longitudinal field, $N_{\textrm{obs}}$ is the number of observations
and the last column indicates whether they were obtained with ESPaDOnS (E) or NARVAL (N).}\label{tab:app} \tabularnewline
  \hline
Name & Night & $B_{z}$ & ${\sigma}_{B_{z}}$ & $N_{\textrm{obs}}$ & E/N\\
   &      &  (G)    & (G) &      & \\
  \hline
\endfirsthead
\multicolumn{6}{c}{\tablename\ \thetable\ -- \textit{Continued from previous page}} \\
  \hline
Name & Night & $B_{z}$ & ${\sigma}_{B_{z}}$ & $N_{\textrm{obs}}$ & E/N\\
   &      &  (G)    & (G) &    & \\
  \hline
\endhead
\hline \multicolumn{6}{r}{\textit{Continued on next page}} \\
\endfoot
\hline
\endlastfoot
$\xi$ Per & 10 Dec. 2006 &            19 &  41 &  3 & N\\
$\xi$ Per & 13 Dec. 2006 &           -10 &  36 &  3 & N\\
$\xi$ Per & 14 Dec. 2006 &            49 &  42 &  3 & N\\
$\xi$ Per & 15 Dec. 2006 &            28 &  22 &  7 & N\\
$\xi$ Per & 16 Dec. 2006 &           -25 & 138 &  1 & N\\
$\xi$ Per & 06 Sep. 2007 &             5 &  25 &  7 & N\\
$\xi$ Per & 07 Sep. 2007 &           -10 &  21 &  6 & N\\
$\xi$ Per & 08 Sep. 2007 &           -20 &  26 &  4 & N\\
$\xi$ Per & 09 Sep. 2007 &            32 &  48 &  2 & N\\
$\xi$ Per & 10 Sep. 2007 &            59 &  55 &  1 & N\\
$\xi$ Per & 11 Sep. 2007 &            46 &  61 &  1 & N\\
$\xi$ Per & 12 Sep. 2007 &            41 &  70 &  1 & N\\
$\xi$ Per & 01 Nov. 2011 &           -15 &  48 &  5 & E\\
  \hline
$\alpha$ Cam & 13 Dec. 2006 &         64 &  35 &  1 & N\\
$\alpha$ Cam & 21 Dec. 2007 &          2 &  10 &  4 & E\\
$\alpha$ Cam & 14 Nov. 2010 &         11 &  20 &  2 & E\\
$\alpha$ Cam & 31 Dec. 2012 &          6 &  24 &  1 & E\\
$\alpha$ Cam & 01 Jan. 2013 &         11 &  25 &  3 & E\\
  \hline
HD 34656 & 11 Nov. 2011 &         -35 &  38 &  1 & E\\
  \hline
$\lambda$ Ori A & 21 Dec. 2007 &         -15 &  23 &  2 & E\\
$\lambda$ Ori A & 18 Jan. 2008 &          36 &  46 &  1 & E\\
$\lambda$ Ori A & 22 Jan. 2008 &         -14 &  22 &  2 & E\\
$\lambda$ Ori A & 14 Oct. 2008 &          31 &  33 &  2 & N\\
$\lambda$ Ori A & 26 Oct. 2008 &          17 &  15 &  7 & N\\
$\lambda$ Ori A & 15 Mar. 2009 &          17 &  30 &  1 & N\\
$\lambda$ Ori A & 17 Mar. 2009 &          12 &  25 &  1 & N\\
$\lambda$ Ori A & 16 Oct. 2010 &          -2 &  12 &  4 & E\\
  \hline
$\epsilon$ Ori & 15 Oct. 2007 &           44 &  75 &  1 & N\\
$\epsilon$ Ori & 17 Oct. 2007 &          -34 &  29 &  6 & N\\
$\epsilon$ Ori & 18 Oct. 2007 &           -3 &   6 & 28 & N\\
$\epsilon$ Ori & 21 Oct. 2007 &            2 &  10 &  8 & N\\
$\epsilon$ Ori & 24 Oct. 2007 &           17 &  13 &  6 & N\\
$\epsilon$ Ori & 13 Oct. 2008 &           20 &   9 &  9 & E\\
$\epsilon$ Ori & 25 Oct. 2008 &           -2 &  13 & 10 & N\\
$\epsilon$ Ori & 15 Mar. 2009 &           26 &  35 &  1 & N\\
$\epsilon$ Ori & 16 Mar. 2009 &            4 &  25 &  1 & N\\
  \hline
15 Mon & 10 Dec. 2006 &            -9 &  50 &  1 & N\\
15 Mon & 15 Dec. 2006 &            -3 &  27 &  1 & N\\
15 Mon & 09 Sep. 2007 &           -17 &  39 &  1 & N\\
15 Mon & 10 Sep. 2007 &            -1 &  30 &  1 & N\\
15 Mon & 11 Sep. 2007 &           -22 &  44 &  1 & N\\
15 Mon & 20 Oct. 2007 &            -2 &  26 &  4 & N\\
15 Mon & 23 Oct. 2007 &            16 &  24 &  4 & N\\
15 Mon & 03 Feb. 2012 &             0 &  20 &  3 & E\\
  \hline
HD~64760 & 21 Nov. 2010 &          51 &  37 &  6 & E\\
HD~64760 & 31 Dec. 2012 &          15 &  59 &  3 & E\\
  \hline
$\zeta$ Pup & 14 Feb. 2012 &         -12 &  21 &  30 & E\\
  \hline
$\zeta$ Oph & 18 Mar. 2011 &          92 &  438 &  1 & N\\
$\zeta$ Oph & 21 Mar. 2011 &        -336 &  361 &  1 & N\\
$\zeta$ Oph & 05 Apr. 2011 &        -266 &  406 &  1 & N\\
$\zeta$ Oph & 08 Jun. 2011 &        -134 &  118 & 20 & E\\
$\zeta$ Oph & 10 Jun. 2011 &         417 &  423 &  1 & N\\
$\zeta$ Oph & 13 Jun. 2011 &        -261 &  426 &  1 & N\\
$\zeta$ Oph & 14 Jun. 2011 &          34 &  310 &  1 & N\\
$\zeta$ Oph & 15 Jun. 2011 &         106 &  377 &  1 & N\\
$\zeta$ Oph & 04 Jul. 2011 &        -206 &  302 &  1 & N\\
$\zeta$ Oph & 07 Jul. 2011 &         127 &  411 &  1 & N\\
$\zeta$ Oph & 10 Jul. 2011 &         170 &  342 &  1 & N\\
$\zeta$ Oph & 11 Jul. 2011 &         -51 &  320 &  1 & N\\
$\zeta$ Oph & 10 Aug. 2011 &          -9 &  315 &  1 & N\\
$\zeta$ Oph & 11 Aug. 2011 &         120 &  345 &  1 & N\\
$\zeta$ Oph & 15 Aug. 2011 &         242 &  374 &  1 & N\\
$\zeta$ Oph & 16 Aug. 2011 &         174 &  335 &  1 & N\\
$\zeta$ Oph & 17 Aug. 2011 &         -85 &  349 &  1 & N\\
$\zeta$ Oph & 18 Aug. 2011 &         -77 &  579 &  1 & N\\
$\zeta$ Oph & 20 Aug. 2011 &         -82 &  418 &  1 & N\\
$\zeta$ Oph & 21 Aug. 2011 &         -91 &  297 &  1 & N\\
$\zeta$ Oph & 22 Aug. 2011 &         108 &  499 &  1 & N\\
$\zeta$ Oph & 23 Aug. 2011 &          14 &  295 &  1 & N\\
$\zeta$ Oph & 26 Aug. 2011 &         455 &  685 &  1 & N\\
$\zeta$ Oph & 27 Aug. 2011 &        -277 &  524 &  1 & N\\
$\zeta$ Oph & 28 Aug. 2011 &         769 &  840 &  1 & N\\
$\zeta$ Oph & 16 Jan. 2012 &        -476 &  446 &  1 & N\\
$\zeta$ Oph & 17 Jan. 2012 &        -444 &  383 &  1 & N\\
$\zeta$ Oph & 24 Jan. 2012 &         -52 &  317 &  1 & N\\
$\zeta$ Oph & 25 Jan. 2012 &          41 &  282 &  1 & N\\
$\zeta$ Oph & 27 Jan. 2012 &        -126 &  596 &  1 & N\\
$\zeta$ Oph & 21 Jun. 2012 &          39 &  351 &  1 & N\\
$\zeta$ Oph & 22 Jun. 2012 &         185 &  316 &  1 & N\\
$\zeta$ Oph & 23 Jun. 2012 &         380 &  298 &  1 & N\\
$\zeta$ Oph & 09 Jul. 2012 &         409 &  348 &  1 & N\\
$\zeta$ Oph & 12 Jul. 2012 &         152 &  369 &  1 & N\\
$\zeta$ Oph & 06 Aug. 2012 &        -132 &  321 &  1 & N\\
$\zeta$ Oph & 07 Aug. 2012 &          75 &  382 &  1 & N\\
$\zeta$ Oph & 08 Aug. 2012 &        -248 &  335 &  1 & N\\
$\zeta$ Oph & 09 Aug. 2012 &         171 &  413 &  1 & N\\
$\zeta$ Oph & 12 Aug. 2012 &        -171 &  491 &  1 & N\\
$\zeta$ Oph & 14 Aug. 2012 &         462 &  361 &  1 & N\\
$\zeta$ Oph & 16 Aug. 2012 &        -161 &  474 &  1 & N\\
$\zeta$ Oph & 17 Aug. 2012 &        -143 &  413 &  1 & N\\
$\zeta$ Oph & 18 Aug. 2012 &        -572 &  448 &  1 & N\\
$\zeta$ Oph & 19 Aug. 2012 &        -178 &  403 &  1 & N\\
$\zeta$ Oph & 20 Aug. 2012 &          56 &  383 &  1 & N\\
  \hline
68 Cyg & 16 Dec. 2006 &           131 &  479 &  1 & N\\
68 Cyg & 10 Sep. 2007 &            80 &  101 &  1 & N\\
68 Cyg & 12 Nov. 2007 &          -106 &  197 &  1 & N\\
68 Cyg & 29 Sep. 2012 &           -11 &   46 &  5 & E\\
  \hline
19 Cep & 13 Dec. 2006 &           -20 &  44 &  1 & N\\
19 Cep & 09 Nov. 2007 &             9 &  59 &  1 & N\\
19 Cep & 13 Nov. 2007 &          -129 &  97 &  1 & N\\
19 Cep & 22 Dec. 2007 &            -8 &  22 &  3 & E\\
19 Cep & 21 Jun. 2008 &           -57 &  28 &  5 & N\\
19 Cep & 22 Jun. 2008 &             9 &  20 &  5 & N\\
19 Cep & 25 Jun. 2008 &             2 &  24 &  3 & N\\
19 Cep & 27 Jun. 2008 &           -18 &  17 &  5 & N\\
19 Cep & 28 Jun. 2008 &            -5 &  19 &  5 & N\\
19 Cep & 26 Jul. 2010 &            23 &  36 &  4 & E\\
  \hline
$\lambda$ Cep & 13 Dec. 2006 &        64 &  79 &  1 & N\\
$\lambda$ Cep & 07 Jul. 2011 &      -100 &  63 &  1 & N\\
$\lambda$ Cep & 08 Jul. 2011 &        36 &  60 &  1 & N\\
$\lambda$ Cep & 10 Aug. 2011 &       -15 &  65 &  1 & N\\
$\lambda$ Cep & 27 Aug. 2011 &       -56 &  83 &  1 & N\\
$\lambda$ Cep & 28 Aug. 2011 &        -2 &  67 &  1 & N\\
$\lambda$ Cep & 16 Jun. 2012 &         4 &  95 &  1 & N\\
$\lambda$ Cep & 22 Jun. 2012 &       -64 &  68 &  1 & N\\
$\lambda$ Cep & 24 Jun. 2012 &        31 &  92 &  1 & N\\
$\lambda$ Cep & 09 Jul. 2012 &       -37 &  78 &  1 & N\\
$\lambda$ Cep & 18 Jul. 2012 &        38 &  63 &  1 & N\\
$\lambda$ Cep & 19 Jul. 2012 &        -9 &  66 &  1 & N\\
$\lambda$ Cep & 22 Jul. 2012 &       -20 &  67 &  1 & N\\
$\lambda$ Cep & 23 Jul. 2012 &        10 &  67 &  1 & N\\
$\lambda$ Cep & 24 Jul. 2012 &         7 &  62 &  1 & N\\
$\lambda$ Cep & 06 Aug. 2012 &        88 &  57 &  1 & N\\
$\lambda$ Cep & 07 Aug. 2012 &      -149 &  80 &  1 & N\\
$\lambda$ Cep & 08 Aug. 2012 &        53 &  67 &  1 & N\\ 
$\lambda$ Cep & 09 Aug. 2012 &        41 &  58 &  1 & N\\
$\lambda$ Cep & 11 Aug. 2012 &       -12 & 126 &  1 & N\\
$\lambda$ Cep & 12 Aug. 2012 &        85 & 103 &  1 & N\\
$\lambda$ Cep & 13 Aug. 2012 &       -66 &  65 &  1 & N\\
$\lambda$ Cep & 15 Aug. 2012 &       -11 &  97 &  1 & N\\
$\lambda$ Cep & 16 Aug. 2012 &        62 &  86 &  1 & N\\
$\lambda$ Cep & 17 Aug. 2012 &       -13 &  86 &  1 & N\\
$\lambda$ Cep & 18 Aug. 2012 &       173 & 149 &  1 & N\\
  \hline
10 Lac & 10 Dec. 2006 &  -6 &   7 &   1 & N\\
10 Lac & 11 Dec. 2006 &  -1 &   7 &   1 & N\\
10 Lac & 13 Dec. 2006 &   8 &   8 &   1 & N\\
10 Lac & 14 Dec. 2006 &  -6 &   7 &   1 & N\\
10 Lac & 15 Dec. 2006 &   8 &   6 &   1 & N\\
10 Lac & 16 Dec. 2006 &  11 &   9 &   1 & N\\
10 Lac & 07 Sep. 2007 &  -2 &   6 &   1 & N\\
10 Lac & 15 Oct. 2007 &   9 &   6 &   3 & N\\
10 Lac & 16 Oct. 2007 &  12 &   6 &   3 & N\\
10 Lac & 17 Oct. 2007 &   1 &   4 &   3 & N\\
10 Lac & 18 Oct. 2007 &  -6 &   8 &   3 & N\\
10 Lac & 19 Oct. 2007 &  -6 &   4 &   3 & N\\
10 Lac & 20 Oct. 2007 &   1 &   4 &   3 & N\\
10 Lac & 21 Oct. 2007 &   5 &   5 &   3 & N\\
10 Lac & 23 Oct. 2007 &   2 &   5 &   3 & N\\
10 Lac & 24 Oct. 2007 &  -3 &   4 &   3 & N\\
10 Lac & 06 Nov. 2007 & -14 &  14 &   1 & N\\
10 Lac & 26 Jul. 2008 &   9 &  10 &   1 & E\\
  \hline
\end{longtable}
\end{center}
\twocolumn

\label{lastpage}

\end{document}